\documentclass[12pt]{article}

\usepackage{PRIMEarxiv}
\usepackage[utf8]{inputenc} % allow utf-8 input
\usepackage{setspace}      
\usepackage[T1]{fontenc}    % use 8-bit T1 fonts
\usepackage{lmodern}  % for better font
\usepackage{hyperref}       % hyperlinks
\usepackage{url}            % simple URL typesetting
\usepackage{booktabs, multicol, multirow}       
\usepackage[table]{xcolor}
\usepackage{natbib}
\usepackage{amsfonts, amsmath, amssymb, amsthm, bm}       %
\usepackage{nicefrac}      
\usepackage{microtype}      % microtypography
\usepackage{lipsum}
\usepackage{fancyhdr}       % header
\usepackage{graphicx}       % graphics
\graphicspath{{media/}}     
\usepackage{algorithm, algpseudocode}

\newcommand{\indep}{\perp \!\!\! \perp}

\newcommand{\A}{{\bf A}}

\newcommand{\B}{{\bf B}}

\newcommand{\D}{{\bf D}}

\newcommand{\mE}{{\mathcal{E}}}

\newcommand{\mG}{{\mathcal{G}}}

\newcommand{\I}{{\bf I}}
\newcommand{\mS}{{\mathcal{S}}}

\newcommand{\T}{{\bf T}}

\newcommand{\W}{{\bf W}}

\newcommand{\X}{{\bf X}}
\newcommand{\x}{{\bm{x}}}
\newcommand{\xx}{{\bm{x}}}

\newcommand{\mZ}{{\mathcal{Z}}}

\newcommand{\bbeta}{\bm{\beta}}

\newcommand{\R}{\mathbb{R}}
\newcommand{\mR}{\mathcal R}
\newcommand{\spc}{{\mS}_{Y|\X}}

\newcommand{\E}{\mathbb{E}}

\newcommand{\Sig}{\bm{\Sigma}}

\newtheorem{theorem}{Theorem}
\newtheorem{lemma}{Lemma}
\newtheorem{proposition}{Proposition}
\newtheorem{definition}{Definition}
\newtheorem{remark}{Remark}

\newtheorem{assumption}{Assumption}

\DeclareMathOperator{\tr}{tr}
\DeclareMathOperator{\ran}{ran}
\DeclareMathOperator{\spn}{span}
\title{Sufficient Dimension Reduction with Multimodal Predictor Objects}

\author{
  Abdul-Nasah Soale \thanks{Corresponding author}\\
  Department of Mathematics, Applied Mathematics, and Statistics, \\
Case Western Reserve University, Cleveland, OH, USA\\
  \texttt{abdul-nasah.soale@case.edu} \\
}

\begin{document}
\maketitle

\begin{abstract}
We study the regression of a scalar response $Y$ on a Euclidean predictor $\X \in \R^p$ and a random predictor object $\mZ$ in a general metric space under arbitrary structural dependence. We introduce a joint sufficient dimension reduction $(\B^\top\X, \mG_Z)$ constructed via a sequential conditional independence framework, which differs fundamentally from the ordinary central subspace of $Y \mid \X$. Methodologically, $\mZ$ is embedded into a reproducing kernel Hilbert space, and a regularized kernel-response envelope model extracts a finite-dimensional reduction of the feature map. Partial adaptive estimation then recovers the Euclidean reduction conditional on the first-stage coordinates. We establish root-$n$ consistency rates for the first-stage envelope subspace under a shrinking regularizer, alongside large-sample consistency for the second-stage Euclidean subspace estimator. Simulations and two real-data applications illustrate the utility and flexibility of the proposed method.
\end{abstract}

% keywords can be removed
\keywords{ Fr\'echet regression \and multimodal data \and nonlinear sufficient dimension reduction \and function-on-linear envelope \and partial dimension reduction }

%\newpage

%==============================================================
\section{Introduction}\label{sec:intro}
%==============================================================
Modern artificial intelligence and complex applications in medicine, agriculture, and engineering often rely on multimodal data featuring non-Euclidean structures, such as images, statistical distributions, temporal functionals, and text \citep{hausdorff1999maturation, petersen2019, chen2023age}. Such non-Euclidean geometry complicates conventional statistical models. Recent advances in Fr\'echet regression \citep{petersen2019, bhattacharjee2023single} and sufficient dimension reduction \citep{ying2022, zhang2024dimension, soale2025detecting} address this by reducing and visualizing random objects while preserving response-relevant information.

While regression with random responses is expanding, regression with multimodal predictor objects remains understudied. With complex multimodal predictor objects, researchers frequently rely on preprocessing pipelines that map each modality into a low-dimensional Euclidean space \citep{staniswalis1998nonparametric, lu2008mpca, verde2015dimension}. However, these unsupervised reductions can discard response-relevant dependence structures and cross-modal associations.

\begin{figure}[ht!]
\centering
\includegraphics[width=0.99\linewidth, ]{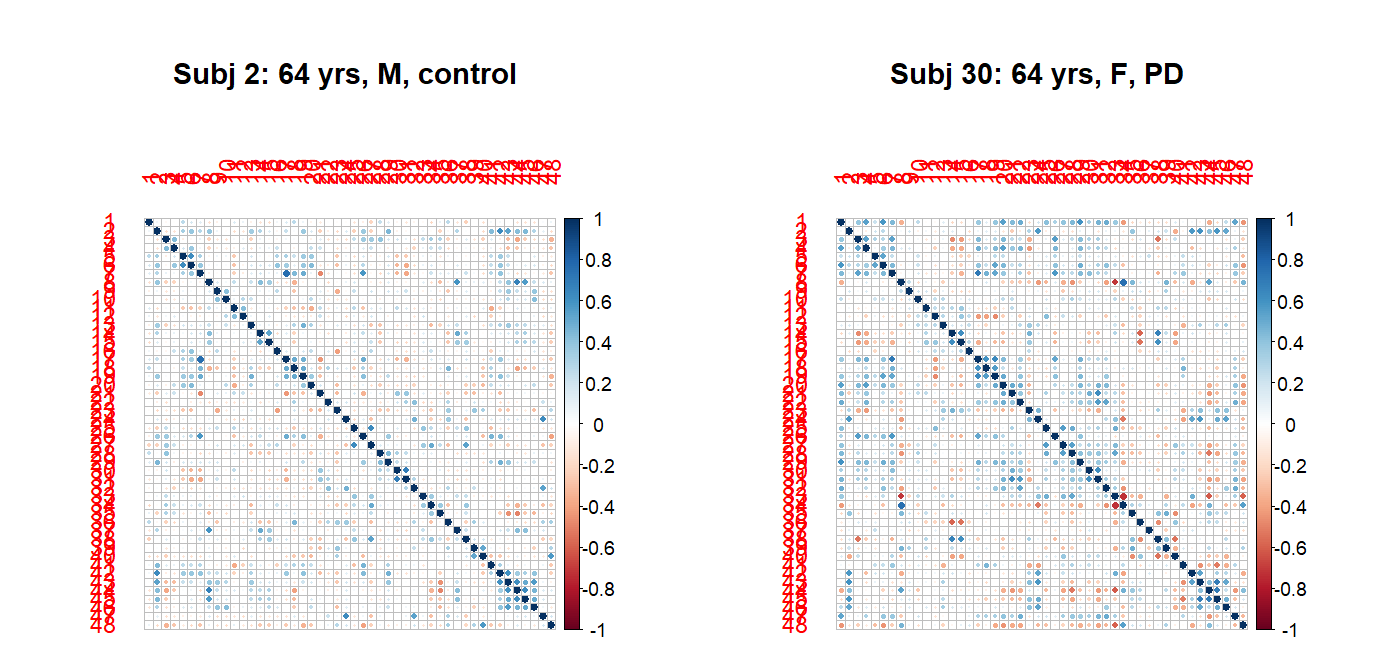}
\caption{Correlation plot between 48 brain regions for a subject with and without Parkinson's disease in the Neurocon dataset of \citet{badea2017exploring}.}
\label{fig:brain_corr}
\end{figure}

For instance, although age and sex are associated with Parkinson's disease \citep{cerri2019parkinson}, adjusting for them when analyzing brain images is difficult. Figure~\ref{fig:brain_corr} illustrates the challenge of separating disease-related regional correlation differences from variation associated with sex. Transforming raw images into anatomical or structural summaries such as gray matter volume \citep{fornito2009mapping}, cortical thickness, or fractional anisotropy \citep{madan2017advances, chen2023age} may improve interpretability, but risks discarding crucial information \citep{usman2026balancing}. This challenge extends to agronomy, where static covariates such as soil texture often need to be modeled jointly with changing meteorological factors like temperature over the growing season \citep{schlenker2009nonlinear}. Other examples can be seen in physiological monitoring, engineering, sports analytics, economics, and chemometrics \citep{matabuena2026multilevel}.

Let $\mZ$ be a non-Euclidean covariate in a metric space $(\Omega_Z,\rho)$, $\X\in\R^p$ a Euclidean covariate, and $Y\in\R$ a scalar response. Because $\X$ and $\mZ$ may be dependent and act nonadditively on $Y$, a joint reduction is required. We adopt the sequential decomposition framework of \citet{yin2015sequential}, treating the different modalities as predictor blocks.

First, $\mZ$ is mapped into a separable Hilbert space via a continuous injection. Such an embedding exists for common spaces, including the 2-Wasserstein space, symmetric positive-definite matrices, spheres, and for temporal functionals under appropriate metrics \citep{zhang2024dimension,soale2026metric}; if $\rho^2$ is of negative type, an isometric embedding exists \citep{zhang2024dimension}. A kernel built on this embedding generates a reproducing kernel Hilbert space feature map for $\mZ$. Given $(\X,Y)$, we reduce $\mZ$ first using an envelope model for the reproducing kernel Hilbert space feature, and then reduce $\X$ via partial sufficient dimension reduction \citep{chiaromonte2002sufficient,feng2013partial}.

Our contributions are as follows. 1) We define a joint sufficient reduction for $(\X,\mZ)$ via conditional independence, show it differs from the ordinary central subspace of $Y\mid\X$, and establish its sequential construction. 2) We propose a function-on-linear response envelope to reduce $\mZ$ without smoothing in $\X$, and a partial minimum average variance estimation approach (MAVE) \citep{xia2002adaptive} using a continuous, estimated conditioning variable to reduce $\X$. 3) We establish consistency rates for the first-stage subspace and consistency for the second-stage estimator under estimated conditioning. 4) We demonstrate via simulations and two applications that this supervised reduction outperforms pipelines based on unsupervised reductions of $\mZ$.

Section~\ref{sec:model} details the framework and embedding of $\mZ$. Section~\ref{sec:sequential_sdr} develops the sequential reduction, while Section~\ref{sec:sample_estimation} discusses the finite-sample estimator and its asymptotic theory. Simulations and real-data applications are presented in Sections~\ref{sec:simulations} and~\ref{sec:application}, respectively. Section~\ref{sec-conc} concludes, and proofs are deferred to the Appendix.

%==============================================================
\section{A joint sufficient reduction framework}\label{sec:model}
%==============================================================
\subsection{Setup}
For a fixed $p \geq 2$, let $\X\in\R^p$, $Y\in\R$, and let $\mZ$ be a random object in a complete separable metric space $(\Omega_Z,\rho)$, all defined on a probability space $(\Omega,\mathcal F,P)$. No restrictions are placed on the dependence between $\X$ and $\mZ$. For random elements $U$, $V$, and $W$, conditional independence $U\indep V\mid W$ means $\E\{h(U)\mid\sigma(V)\vee\sigma(W)\}=\E\{h(U)\mid\sigma(W)\}$ almost surely for every bounded measurable function $h$, where $\sigma(W)$ denotes the $\sigma$-field generated by $W$. Because all spaces are Polish, the regular conditional distributions $P_{Y|(\mZ,\X)}$ and $P_{\mZ|(Y,\X)}$ exist. Throughout, $\|\cdot\|$ denotes the operator norm, $\|\cdot\|_{\mathrm{HS}}$ is the Hilbert--Schmidt norm, and $\operatorname{ker}(A)$, $\operatorname{ran}(A)$, and $A^*$ denote the null space, range, and adjoint of an operator $A$.

\subsection{Sufficient reduction}
\begin{definition}[Central $\sigma$-field of $\mZ$ given $\X$]\label{def:G}
A sub $\sigma$-field $\mG\subseteq\sigma(\mZ)$ is \emph{sufficient} if
\begin{equation}\label{eq:csigmafield}
Y\indep\mZ\mid(\X,\mG).
\end{equation}
If a sufficient $\mG_Z$ is contained in every sufficient sub $\sigma$-field, it is the \emph{central $\sigma$-field of $\mZ$ for $Y$ given $\X$}.
\end{definition}

\begin{definition}[Partial central subspace given $\mG$]\label{def:S}
For a sub $\sigma$-field $\mG\subseteq\sigma(\mZ)$, a subspace $\mS\subseteq\R^p$ is a \emph{dimension reduction subspace given $\mG$} if
\begin{equation}\label{eq:partialcs}
Y\indep\X\mid(P_{\mS}\X,\mG).
\end{equation}
The smallest such subspace, when it exists, is the \emph{partial central subspace} $\mS^{(\mG)}_{Y|\X}$. We write $\spn(\B)=\mS^{(\mG)}_{Y|\X}$ with $\B\in\R^{p\times d_x}$,
$\B^\top\B=I_{d_x}$, $d_x<p$.
\end{definition}

\begin{assumption}\label{ass:exist}
The conditional distributions $P_{\mZ|(\X,Y)}$ and $P_{Y|(\X,\mZ)}$ are dominated by a $\sigma$-finite measure, the central $\sigma$-field $\mG_Z$ exists, and
$\mS^{(\mG_Z)}_{Y|\X}$ exists.
\end{assumption}
\noindent Assumption~\ref{ass:exist} extends the dominating-measure condition of Theorem~1 in \citet{lee2013general}, which concerns $Y\mid\mZ$ without an additional conditioning variable. For the subspace, existence follows under the conditions of \citet{yin2008successive}.

\begin{theorem}\label{thm:joint}
Let $\mG \subseteq \sigma(\mZ)$ satisfy \eqref{eq:csigmafield} and let $\spn(\B)$ be a conditional dimension reduction subspace given $\mG$, satisfying \eqref{eq:partialcs}. Then, 
\begin{equation}\label{eq:joint_independence}
Y \mathrel{\indep} (\X, \mZ) \mid (\B^\top\X, \mG).
\end{equation}
\end{theorem}

\begin{remark}\label{rem:examples}
(a) {\it Conditioning on $\X$ matters.} Let $Y$ and $\mZ$ be independent $\text{Bernoulli}(1/2)$ variables and $X = \mZ \oplus Y$. Then $Y \indep \mZ$ marginally, but $Y = X \oplus \mZ$ is deterministic given $X$. Thus, conditioning on $X$ makes $\mZ$ informative.

(b) {\it $\B$ differs from the central subspace of $Y\mid\X$.} Let $X_1, X_2, \eta, \epsilon$ be independent, $\mZ=X_2+\eta$, and $Y=X_1+\mZ+\epsilon$. Then $\mG=\sigma(\mZ)$ and $\spn(\B)=\spn(e_1)$, whereas the central subspace of $Y$ given $\X$ is $\spn\{(1,1)^\top\}$.

(c) {\it The regression model does not uniquely identify $\B$.} Let $\X = (X_1, X_2)^\top$ have independent components, $\mZ = X_2$, and $Y = X_1 + \mZ + \epsilon$ with $\epsilon \indep \X$. The model $Y = g(\B^\top\X, \mZ) + \epsilon$ is satisfied by both $\B = e_1$ and $\B = (1,1)^\top$. Because $\spn(e_1) \cap \spn\{(1,1)^\top\} = \{\mathbf{0}\}$, a unique minimal conditional central subspace does not exist.

(d) {\it Sufficiency versus minimality.} Theorem~\ref{thm:joint} guarantees sufficiency but not minimality. Let $\mZ \in \R$ be independent of $\X$, and $Y = \varphi(X_1 + \mZ) + \epsilon$ with $\epsilon \indep (\X, \mZ)$. Then $X_1 + \mZ$ is minimal sufficient for $Y \mid (\X, \mZ)$. For $\mG = \sigma(\mZ)$, Theorem~\ref{thm:joint} yields the sufficient but non-minimal joint reduction $(X_1, \sigma(\mZ))$ since $\sigma(X_1 + \mZ) \subsetneq \sigma(X_1) \vee \sigma(\mZ)$.
\end{remark}

\subsection{Embedding and kernel for $\mZ$}\label{subsec:embed}
\begin{assumption}\label{ass:embed}
There exists a separable Hilbert space $\mathcal H$ and a continuous injection $\Phi:\Omega_Z\to\mathcal H$.
\end{assumption}
This is adopted from \citealp[Prop.~2]{zhang2024dimension}. It holds for the 2-Wasserstein space of probability distributions, for symmetric positive definite (SPD) matrices under the Frobenius metric, for the sphere with the geodesic metric, and trivially when $\Omega_Z$ is a separable Hilbert space such as $L^2(\mathbb I)$. Following the nested Hilbert-space construction of \citet{lee2022functional} and \citet{sang2026nonlinear}, define
\begin{equation}\label{eq:kernel}
\kappa(\mathbf z,\mathbf z')=\varsigma\big(\|\Phi(\mathbf z)\|^2,\langle\Phi(\mathbf z),\Phi(\mathbf z')\rangle,\|\Phi(\mathbf z')\|^2\big)
\end{equation}
for a function $\varsigma$ making $\kappa$ positive definite. The standard choice is the Gaussian-type kernel
$\kappa(\mathbf z,\mathbf z')=\exp\{-\gamma\|\Phi(\mathbf z)-\Phi(\mathbf z')\|^2\}$ but the Laplace-type kernel $\exp\{-\gamma\|\Phi(\mathbf z)-\Phi(\mathbf z')\|\}$ is also admissible. Polynomial and linear kernels, which depend on $\langle\Phi(\mathbf z),\Phi(\mathbf z')\rangle$ only, are not recommended because they fail Assumption~\ref{ass:kernel} \citep{sang2026nonlinear}. On the sphere with geodesic distance, the Gaussian-type kernel is not positive definite \citep{feragen2015geodesic} whereas the Laplace-type kernel is.

Let $\mathfrak M_Z$ be the RKHS of $\kappa$ and $\Phi(\mathbf z)=\kappa(\cdot,\mathbf z)$ the feature map.

\begin{assumption}\label{ass:kernel}
(a) $\kappa$ is continuous, bounded by $\bar\kappa<\infty$, and $\Phi$ is injective. (b) $\mathfrak M_Z+\R$ is dense in $L_2(P_Z)$.
\end{assumption}
Boundedness implies $\E [f^2(\mZ)] \le\bar\kappa\|f\|^2_{\mathfrak M_Z}$ for $f\in\mathfrak M_Z$. Thus, the inclusion $\mathfrak M_Z\to L_2(P_Z)$ is bounded. A sufficient condition for (b) is that
$\kappa$ be a bounded cc-universal kernel \citep[Prop.~2, Thm.~1]{zhang2024dimension}; see also \citet{micchelli2006universal,sriperumbudur2011universality}. For the Gaussian-type
kernel with $\gamma>0$, this holds under Assumption~\ref{ass:embed}.

\begin{lemma}\label{lem:sigma}
Under Assumptions~\ref{ass:embed} and~\ref{ass:kernel}(a), $\Phi$ is a Borel isomorphism onto its range and $\sigma\{\Phi(\mZ)\}=\sigma(\mZ)$.
\end{lemma}
Lemma~\ref{lem:sigma} follows directly by the Lusin--Souslin theorem \citep{cohn2013measure}. Assumption~\ref{ass:kernel}(b) ensures that linear functionals of $\Phi(\mZ)$ approximate any square-integrable function of $\mZ$. Therefore, a model that is linear in the feature map $\Phi$ is not restrictive in principle, as it retains the capacity to approximate arbitrary non-linear functional relationships in the original data space. 

%==============================================================
\section{Sequential reduction approach}\label{sec:sequential_sdr}
%==============================================================
The target in Section~\ref{sec:model} is defined sequentially. First $\mZ$ given $\X$, then $\X$ given the first-stage $\sigma$-field. We adopt the sequential decomposition idea of
\citet{yin2015sequential} in a different role. In their formulation, they split a high-dimensional $\X$ into blocks to avoid inverting a large covariance matrix, whereas here the blocks are predictors in different spaces and the decomposition follows from Theorem~\ref{thm:joint}.

\begin{lemma}[Prop.~1(a) of \citealp{yin2015sequential}]\label{lem:seq}
Let $T$ be a measurable function of $\Phi(\mZ)$ with $\Phi(\mZ)\indep(\X,Y)\mid T$. Then $\mZ \indep Y\mid(\X,\sigma(T))$, i.e.\ $\sigma(T)$ satisfies \eqref{eq:csigmafield}.
\end{lemma}

\subsection{Response-envelope reduction of $\mZ$}\label{subsec:response-envelope}
Treating $(Y,\X)$ as a joint predictor of $\mZ$, we employ the parametric inverse regression approach of \citet{bura2001estimating} within a response envelope formulation \citep{cook2010envelope,su2023envelope}. Let $\mathbf f(Y)=(f_1(Y),\dots,f_r(Y))^\top$ be bounded measurable transformations of $Y$, where $f_1$ is chosen to be injective (such as identity, slice indicators or B-splines). Since $f_1$ is an injective Borel map on a Polish space, it admits a Borel inverse by the Lusin--Souslin theorem, ensuring that $\sigma\{\mathbf f(Y)\}=\sigma(Y)$. Let $\W=(\mathbf f(Y)^\top,\X^\top)^\top\in\R^{q}$ with $q=r+p$, and define the centered variables $\widetilde\Phi=\Phi(\mZ)-\mu_Z$ and $\widetilde\W=\W-\E(\W)$, where $\mu_Z=\E[\Phi(\mZ)]$. 

By Assumption~\ref{ass:kernel}(a), $\|\widetilde\Phi\|\le2\sqrt{\bar\kappa}$. It follows that $\E\|\widetilde\Phi\|^2 \le 4\bar\kappa < \infty$, which guarantees that the covariance operator $\Sigma_Z = \E(\widetilde\Phi\otimes\widetilde\Phi)$ is trace-class and well-defined, where the tensor product operator is defined as $(a\otimes b)h=\langle b,h\rangle_{\mathfrak M_Z}a$. We assume that $\E\|\W\|^2<\infty$ and that $\Sigma_W=\E(\widetilde\W\widetilde\W^\top)$ is nonsingular. Finally, we define the cross-covariance operator $\Sigma_{ZW}=\E(\widetilde\Phi\,\widetilde\W^\top)$. Note that because $\Sigma_Z$ is a compact trace-class operator, its inverse $\Sigma_Z^{-1}$ is unbounded, which motivates the low-dimensional envelope structure.

The Hilbert-space linear projection of $\widetilde\Phi$ on $\widetilde\W$ is given by
\begin{equation}\label{eq:proj}
\widetilde\Phi = \bbeta\,\widetilde\W + \varepsilon, \quad \bbeta = \Sigma_{ZW}\Sigma_W^{-1}, \quad \E(\varepsilon) = 0, \quad \E(\varepsilon \otimes \widetilde\W) = \mathbf{0},
\end{equation}
where $\bbeta : \R^q \to \mathfrak M_Z$ is a finite-rank linear operator. Equation~\eqref{eq:proj} defines a geometric population projection rather than a structural model. As such, it always exists and requires neither the linearity of $\E(\Phi \mid \W)$ nor the Gaussianity of the errors. The corresponding residual covariance operator $\Sigma_\varepsilon = \E(\varepsilon \otimes \varepsilon)$ is given by
\begin{equation}\label{eq:residual-covariance}
\Sigma_\varepsilon = \Sigma_Z - \bbeta\Sigma_W\bbeta^* = \Sigma_Z - \Sigma_{ZW}\Sigma_W^{-1}\Sigma_{WZ},
\end{equation}
where $\bbeta^*$ is the Hilbert-space adjoint of $\bbeta$, and $\Sigma_{WZ} = \Sigma_{ZW}^* : \mathfrak M_Z \to \R^q$ is the adjoint of the cross-covariance operator. Linearity in the feature map $\Phi$ still accommodates fully non-linear dependence on $\mZ$ and, through $\mathbf f(Y)$, non-linear dependence on $Y$. If greater flexibility in the $\X$ space is required, $\X$ in $\W$ can be replaced by a non-linear basis expansion $\mathbf h(\X)$ satisfying $\sigma\{\mathbf h(\X)\} = \sigma(\X)$, which remains computationally and theoretically feasible since the dimension $p$ is fixed.

\begin{definition}
A closed subspace $\mathcal S \subseteq \mathfrak M_Z$ \emph{reduces} $\Sigma_\varepsilon$ if both $\mathcal S$ and $\mathcal S^\perp$ are invariant under $\Sigma_\varepsilon$. The $\Sigma_\varepsilon$-\emph{envelope} of $\ran\bbeta$, denoted by $\mE = \mE_{\Sigma_\varepsilon}(\ran\bbeta)$, is the intersection of all reducing subspaces of $\Sigma_\varepsilon$ that contain $\ran\bbeta$ \citep{su2023envelope}.   
\end{definition}

Let $P_\mE$ be the orthogonal projection operator onto $\mE$ and $Q_\mE = I - P_\mE$. By construction, the following properties hold:
\begin{equation}\label{eq:envelope-properties}
\ran\bbeta \subseteq \mE, \qquad Q_\mE\bbeta = \mathbf{0}, \qquad P_\mE\Sigma_\varepsilon Q_\mE = Q_\mE\Sigma_\varepsilon P_\mE = \mathbf{0},
\end{equation}
where $\mathbf{0}$ denotes the zero operator. We define $U = P_\mE\widetilde\Phi$ and $V = Q_\mE\widetilde\Phi$ as the \emph{material} and \emph{immaterial} parts of $\widetilde\Phi$, respectively. Applying the structural projection \eqref{eq:proj} alongside the envelope properties in \eqref{eq:envelope-properties}, the immaterial part completely decouples from $\widetilde\W$, yielding
\begin{equation}\label{eq:UV}
U = \bbeta\widetilde\W + P_\mE\varepsilon, \qquad V = Q_\mE\varepsilon.
\end{equation}

\begin{assumption}\label{ass:sparse}
$\ran\bbeta$ is contained within a finite sum of eigenspaces of $\Sigma_\varepsilon$. Consequently, the envelope $\mE$, which is spanned by the projections of $\ran\bbeta$ onto these eigenspaces, has a finite dimension $\dim(\mE) = u < \infty$.
\end{assumption}

This formulation is adapted from \citet[Def.~1(ii)]{su2023envelope} and represents a genuine restriction in the infinite-dimensional setting. Without it, the envelope constructed as a sum of eigenspaces of the compact operator $\Sigma_\varepsilon$ could be infinite-dimensional. Note that if any eigenvalues of $\Sigma_\varepsilon$ are repeated, $\mE$ may form a proper subspace of the eigenspaces it intersects.

\begin{assumption}\label{ass:cond_indep}
The immaterial part $V$ is conditionally independent of the predictors given the material part $U$, denoted as $V \indep \W \mid U$, or equivalently, $Q_\mE\varepsilon \indep \W \mid (\bbeta\widetilde\W + P_\mE\varepsilon)$.
\end{assumption}

The covariance orthogonality in \eqref{eq:envelope-properties} does not imply the conditional independence in Assumption~\ref{ass:cond_indep}, which drives the subsequent sufficiency results. While it holds trivially when $\mE = \mathfrak M_Z$, the envelope framework aims to identify a low-dimensional subspace satisfying this assumption. It can fail if the conditional distribution of the immaterial residual $Q_\mE\varepsilon$ depends on $\W$ beyond its projection onto $U$ (e.g., under heteroscedasticity) or if its conditional expectation depends nonlinearly on $\W$. It holds, however, under the following structural cases.

\begin{proposition}\label{prop:cond_sufficient}
Assumption~\ref{ass:cond_indep} holds if either of the following conditions is satisfied\\
(i) $Q_\mE\varepsilon \indep (P_\mE\varepsilon, \widetilde\W)$, or 
(ii) $\varepsilon$ is Gaussian and $\varepsilon \indep \widetilde\W$.
\end{proposition}

Case (ii) cannot occur for a bounded kernel with a nondegenerate conditional error $\varepsilon \mid \widetilde\W$, as the resulting feature vector $\widetilde\Phi = \bbeta\widetilde\W + \varepsilon$ would have unbounded support, contradicting $\|\widetilde\Phi\| \le 2\sqrt{\bar\kappa}$. Nonetheless, Case (ii) is highly relevant for linear kernels over infinite-dimensional Hilbert coordinates. For bounded kernels, the conditional independence assumption must instead be justified via path (i) or verified directly through empirical diagnostics (see Remark~\ref{rem:diagnostic}).

\begin{theorem}\label{thm:suff}
Suppose Assumptions~\ref{ass:embed}, \ref{ass:kernel}(a), \ref{ass:sparse}, and \ref{ass:cond_indep} hold. Then
\begin{equation}\label{eq:feature-sufficiency}
\widetilde\Phi \indep \W \mid U .
\end{equation}
Moreover, let $\Gamma : \R^u \to \mathfrak M_Z$ be a linear isometry with range $\mE$, and define the finite-dimensional coordinate vector $T = \Gamma^*\widetilde\Phi \in \R^u$. Then the sub $\sigma$-field $\mG_1 = \sigma(T) \subseteq \sigma(\mZ)$ satisfies \eqref{eq:csigmafield}, implying that
\begin{equation}\label{eq:response-envelope-sufficiency}
Y \indep \mZ \mid (\X, T).
\end{equation}
\end{theorem}

\begin{remark}\label{rem:minimal}
By definition, the conditional central $\sigma$-field satisfies $\mG_Z \subseteq \mG_1 = \sigma(T)$. Thus, the first-stage envelope reduction is sufficient but not necessarily minimal. Theorem~\ref{thm:joint} applies with $\mG = \mG_1$, though the associated dimension reduction subspace $\B(\mG_1)$ estimated in Stage~2 may differ from $\B(\mG_Z)$. These two subspaces coincide if and only if the component of $\mG_1$ orthogonal to $\mG_Z$ is independent of $(\X, Y)$, which holds in our numerical simulations.
\end{remark}

\begin{remark}\label{rem:diagnostic}
Assumption~\ref{ass:cond_indep} can be evaluated empirically from the model fit. One can compute the estimated immaterial scores $\widehat{Q}_\mE\widetilde\Phi_i$, remove their nonlinear dependence on the material coordinates $\widehat{T}_i$ via nonparametric kernel regression, and test the independence of the residualized scores from $\W_i$ using a distance-covariance test or the Hilbert--Schmidt independence criterion \citep{szekely2007}.
\end{remark}

{\bf Regularized population objective.} Following \citet{cook2016note}, the envelope subspace $\mE$ in a finite-dimensional framework can be characterized as the minimizing subspace of the objective function
\begin{equation}\label{eq:Jlam0}
J(\mG)=\log\big|\Gamma^*\Sigma_\varepsilon\Gamma\big|+\log\big|\Gamma^*\Sigma_Z^{-1}\Gamma\big|,
\end{equation}
for a $u$-dimensional subspace $\mG$ with an orthonormal basis operator $\Gamma: \R^u \to \mathfrak M_Z$ \citep[see][Lemma A.15]{cook2018introduction}. However, because $\Sigma_Z$ is a trace-class operator in our infinite-dimensional Hilbert setting, its inverse $\Sigma_Z^{-1}$ is unbounded, rendering $\Gamma^*\Sigma_Z^{-1}\Gamma$ generally undefined and its log-determinant structurally invalid. Moreover, while the injectivity of $\Sigma_\varepsilon$ guarantees that $\Gamma^*\Sigma_\varepsilon\Gamma$ is positive definite, it does not ensure that the subsequent sample optimization is numerically stable. Therefore, we introduce regularization to overcome these theoretical and computational challenges.

For regularization parameters $\lambda, \lambda_1 >0$ with $\lambda_1 \ll \lambda$, we define the regularized population objective function as
\begin{equation}\label{eq:Jlam}
J_{\lambda, \lambda_1}(\mG)=\log\big|\Gamma^*(\Sigma_\varepsilon + \lambda_1 I)\Gamma\big|+\log\big|\Gamma^*(\Sigma_Z+\lambda I)^{-1}\Gamma\big|,
\end{equation}
where $I$ is the identity operator on $\mathfrak M_Z$. Crucially, because $\Gamma$ maps from $\R^u$ into the Hilbert space, the arguments inside the log-determinants in \eqref{eq:Jlam} reduce to standard $u \times u$ symmetric, positive-definite matrices rather than infinite-dimensional operators. The objective $J_{\lambda, \lambda_1}$ is invariant to the specific choice of basis, as any orthogonal transformation $\Gamma \mapsto \Gamma O$ (where $O \in \R^{u \times u}$ is an orthogonal matrix) leaves both matrix determinants unchanged. 

To reduce the computational burden of tuning two separate hyper-parameters, one may set $\lambda_1 = \lambda$, which yields the single-parameter objective $J_\lambda(\mG) \equiv J_{\lambda, \lambda}(\mG)$. We adopt this single-parameter convention when deriving our asymptotic theoretical properties. Under Assumption~\ref{ass:sparse} and Assumption~\ref{ass:S1}(d) below, the true envelope $\mE$ remains the unique minimizer of $J_\lambda(\mG)$ over all $u$-dimensional subspaces for every $\lambda>0$. This regularization introduces no population-level bias, and its sample counterpart serves purely to make the empirical optimization well-posed. In settings where $\lambda_1 \ll \lambda$, the corresponding theoretical guarantees continue to hold by replacing $\lambda$ with $\min(\lambda_1, \lambda)$.

\subsection{Stage 2: Partial OPG and MAVE Given $T$}\label{subsec:partial}
We now condition on the sub $\sigma$-field $\mG_1=\sigma(T)$ and denote the conditional dimension reduction subspace as $\mS=\mS^{(\mG_1)}_{Y|\X}=\spn(\B)$. For each component $k\le r$, let $m_k(\xx,\mathbf t)=\E\{f_k(Y)\mid\X=\xx, T=\mathbf t\}$ represent the conditional mean function, and let $\mS_k$ denote the smallest subspace satisfying the dimension reduction constraint $m_k(\xx,\mathbf t)=m_k(P_{\mS_k}\xx,\mathbf t)$. We define the partial outer product of gradient matrices as
\[
M^{(k)}=\E\big\{\nabla_{\xx}m_k(\X, T)\,\nabla_{\xx}m_k(\X, T)^\top\big\},\qquad M=\sum_{k=1}^rM^{(k)},
\]
where the expectation is taken over the joint distribution of $(\X, T)$.

\begin{proposition}\label{prop:opg}
(a) If $m_k(\xx, \mathbf{t})$ is continuously differentiable in $\xx$ with $\E\|\nabla_\xx m_k(\X,T)\|^2<\infty$, then $\spn(M^{(k)})\subseteq\mS_k\subseteq\mS$. Furthermore, if for each fixed $\mathbf t$ the conditional support of $\X$ given $T=\mathbf t$ is open and convex, then $\spn(M^{(k)})=\mS_k$.

(b) Define $\mS':=\spn(\cup_{k\le r}\mS_k)$. Then $\mS'\subseteq\mS$ for every finite $r$. If the family of transformations $\{f_k\}_{k\ge1}$ spans a dense subset of $L^2(P_Y)$, then $\spn(\cup_{k\ge1}\mS_k)=\mS$.
\end{proposition}

Proposition \ref{prop:opg} adapts the classic non-parametric gradient estimation properties of \citet{xia2002adaptive} and the sequential rules of \citet{yin2015sequential} to our specific conditioning environment. Note that when utilizing a finite number of transformations $r$, only containment ($\mS' \subseteq \mS$) is guaranteed. Achieving full exhaustiveness theoretically requires a rich enough family of transformations, which can be accomplished in practice by allowing $r$ to grow adaptively with the sample size.

\begin{remark}[Mean-based variant]
Taking $r=1$, $f_1(y)=y$ gives the partial central \emph{mean} subspace, which may differ from the central space. For the central mean space, Sections~\ref{sec:model} and the sufficiency claim then need the central mean $\sigma$-field and subspace.
\end{remark}

%==============================================================
\section{Estimation and theoretical properties}\label{sec:sample_estimation}
%==============================================================
\subsection{Estimation in Finite Samples}\label{subsec:estimation}
Let $\{(\X_i,Y_i,\mZ_i)\}_{i=1}^n$ be an i.i.d. sample drawn from the joint distribution of $(\X,Y,\mZ)$.

{\bf Stage 1: Kernel Response Envelope.}
Let $K = (\kappa(\mZ_i,\mZ_j))_{ij} \in \R^{n \times n}$ denote the Gram matrix, and define the centering matrix $H = I_n - \mathbf 1_n\mathbf 1_n^\top/n$ and the centered Gram matrix $G = HKH$. Let $\widehat\mu_Z = n^{-1}\sum_{i=1}^n \kappa(\cdot,\mZ_i)$, and define the operator $\Psi : \R^n \to \mathfrak M_Z$ via $\Psi\mathbf a = \sum_{i=1}^n a_i \{\kappa(\cdot,\mZ_i) - \widehat\mu_Z\}$. This construction ensures that $\Psi^*\Psi = G$ and that the sample covariance operator is given by $\widehat\Sigma_Z = n^{-1}\Psi\Psi^*$. 

Let $\widetilde\W \in \R^{n \times q}$ be the matrix of centered predictors, and let $\Pi_W = \widetilde\W(\widetilde\W^\top\widetilde\W)^{-1}\widetilde\W^\top$ denote its associated projection matrix. The ordinary least squares (OLS) fit yields the sample regression operator $\widehat\bbeta = \Psi\widetilde\W(\widetilde\W^\top\widetilde\W)^{-1}$ and the sample residual covariance operator $\widehat\Sigma_\varepsilon = n^{-1}\Psi(I_n - \Pi_W)\Psi^* = \widehat\Sigma_Z - \widehat\bbeta\widehat\Sigma_W\widehat\bbeta^*$. No regularization is required for the estimation of $\widehat\bbeta$ because $\Sigma_W$ is a finite $q \times q$ matrix assumed to be invertible. 

Since all sample operators have their ranges contained within the finite-dimensional subspace $\ran(\Psi)$, we can represent the basis operator of the regularized envelope as $\Gamma = \Psi A$, where $A \in \R^{n \times u}$. Utilizing the operator identity $(\widehat\Sigma_Z + \lambda I)^{-1}\Psi = \Psi(G/n + \lambda I_n)^{-1}$, the sample counterpart to the regularized population objective function \eqref{eq:Jlam} (with $\lambda_1 = \lambda$) evaluates to
\begin{equation}\label{eq:Jn}
\widehat J_\lambda(A) = \log|A^\top M_\varepsilon A| + \log|A^\top M_Z A| - 2\log|A^\top GA|,
\end{equation}
where the $n \times n$ coordinate matrices are defined symmetrically as
\begin{equation}\label{eq:M_matrices}
M_\varepsilon = n^{-1}G(I_n - \Pi_W)G + \lambda G, \qquad M_Z = n G(G + n\lambda I_n)^{-1}G.
\end{equation}
The subtraction of the final log-determinant term in \eqref{eq:Jn} ensures that $\widehat J_\lambda$ is invariant to the choice of basis transformation $A \mapsto AO$ for any invertible matrix $O \in \R^{u \times u}$. Consequently, $\widehat J_\lambda$ is a well-defined objective function on the Grassmannian manifold $\mathbb{G}(u, n)$, and the standard scale normalization $A^\top GA = I_u$ can be imposed post-optimization. 

Upon minimizing \eqref{eq:Jn} to obtain the estimator $\widehat{A}$, the finite-dimensional coordinate vectors for the training observations are given by the columns of the $u \times n$ matrix $\widehat T = \widehat{A}^\top G$. For a newly observed conditioning state $\mathbf z \in \Omega_Z$, its estimated material coordinate vector is computed as:
\begin{equation}\label{eq:new_score}
\widehat T(\mathbf z) = \widehat A^\top H \big(\mathbf k_{\mathbf z} - n^{-1}K\mathbf 1_n\big), \quad \text{where } \mathbf k_{\mathbf z} = (\kappa(\mZ_i, \mathbf z))_i \in \R^n.
\end{equation}
We carry out the minimization of \eqref{eq:Jn} directly on the Grassmannian manifold. Although $\widehat\Gamma$ is identified only up to an arbitrary orthogonal rotation, the subsequent Stage~2 estimation procedures respect this rotational indeterminacy through the use of isotropic kernels over the coordinate space of $T$.

{\bf Stage 2: Partial OPG.}
We first standardize the predictors $\X_i$ and the estimated material coordinates $\widehat T_i$. For each focal observation $j\le n$ and each transformation component $k\le r$, we solve the localized linear regression problem by minimizing the following objective over $(a,\mathbf b,\mathbf c)\in\R\times\R^p\times\R^u$:
\begin{equation}\label{eq:opg_local}
\sum_{i=1}^n w_{ij}\big\{f_k(Y_i)-a-\mathbf b^\top(\X_i-\X_j)-\mathbf c^\top(\widehat T_i-\widehat T_j)\big\}^2,
\end{equation}
where the weights $w_{ij}$ are constructed using a product of isotropic kernels normalized such that $\sum_{i=1}^n w_{ij}=1$:
\[
w_{ij} = K_x\Big(\tfrac{\|\X_i-\X_j\|}{h_x}\Big)K_t\Big(\tfrac{\|\widehat T_i-\widehat T_j\|}{h_t}\Big)/\displaystyle\sum_{m=1}^n K_x\Big(\tfrac{\|\X_m-\X_j\|}{h_x}\Big)K_t\Big(\tfrac{\|\widehat T_m-\widehat T_j\|}{h_t}\Big).
\]
Let $\widehat{\mathbf b}_{jk}\in\R^p$ denote the minimizing slope vector with respect to $\X$, which serves as a local estimator for the gradient $\nabla_{\xx}m_k(\X_j, \widehat T_j)$. We then construct the sample partial OPG matrix:
\[
\widehat M = \frac{1}{n}\sum_{j=1}^n \sum_{k=1}^r \widehat{\mathbf b}_{jk}\widehat{\mathbf b}_{jk}^\top.
\]
The subspace spanned by the top $d_x$ eigenvectors of $\widehat M$ forms our final estimator for the dimension reduction subspace $\mS = \spn(\B)$. 

Note that the use of a product kernel structure does not require the independence of $\X$ and $\mZ$ as any structural dependence between them is naturally accommodated through the joint density of $(\X,T)$. Because the local smoothing operates simultaneously across $p+u$ dimensions, we initialize the bandwidths symmetrically as $h_x = h_t = h$, where the baseline starting bandwidth is given by $h = c\,n^{-1/(p_0+6)}$ with $p_0 = \max(p+u, 3)$ following the adaptive rate considerations in \citet{xia2002adaptive}.

{\bf Stage 2: Partial MAVE (Refinement).}
Given a current estimate $\widehat\B\in\R^{p\times d}$ satisfying $\widehat\B^\top\widehat\B=I_d$, we refine the subspace by updating the localized weights as a function of $\B$:
\[
w_{ij}(\B) = K_x\Big(\tfrac{\|\B^\top(\X_i-\X_j)\|}{h_x}\Big)K_t\Big(\tfrac{\|\widehat T_i-\widehat T_j\|}{h_t}\Big) / \sum_{m=1}^n K_x\Big(\tfrac{\|\B^\top(\X_m-\X_j)\|}{h_x}\Big)K_t\Big(\tfrac{\|\widehat T_m-\widehat T_j\|}{h_t}\Big).
\]
The updated subspace estimator is then defined as the global minimizer of the localized average variance:
\begin{equation}\label{eq:mave_objective}
\widehat\B = \arg\min_{\B^\top\B=I_d}\sum_{k=1}^r \sum_{j=1}^n \min_{a, \mathbf d, \mathbf c} \sum_{i=1}^n w_{ij}(\B)\big\{f_k(Y_i)-a-\mathbf d^\top\B^\top(\X_i-\X_j)-\mathbf c^\top(\widehat T_i-\widehat T_j)\big\}^2,
\end{equation}
where $\mathbf{d} \in \R^d$ represents the local coordinate slope vector. We solve \eqref{eq:mave_objective} by alternating iteratively between optimizing the local parameters $(a,\mathbf d,\mathbf c)$ for a fixed $\B$ and updating the global basis matrix $\B$ for fixed local parameters. The routine is initialized using the partial OPG estimator from the previous step, with the bandwidths adaptively updated at each pass. The structural dimension $d_x$ is selected via cross-validation or predictor augmentation \citep{luo2021order}, though we note that the theoretical validity of the latter when conditioned on an estimated score $\widehat T$ has not yet been formally established.

{\bf Tuning.}
%The response transformation family $\mathbf f(Y)$ is implemented using $r$ slice indicators or a collection of B-splines. 
We initialize the kernel bandwidth parameter by setting $\gamma = 1/(2\sigma^2)$, where $\sigma^2$ represents the mean pairwise squared distance evaluated over the sample observations, and subsequently refine it during optimization. The hyper-parameter triplet $(u, \lambda, \gamma)$ is selected via $K$-fold cross-validation nested within the Stage-2 fit. The optimization criterion evaluates the $K$-fold squared prediction error of a non-parametric smoother of $Y$ regressed on the joint vector $(\widehat\B^\top\X, \widehat T)$. We explicitly note that the BIC and rank-based tests developed by \citet{su2023envelope} are exact only for unregularized, finite-dimensional models, and are thus omitted here. Bandwidths are initialized as $h_x = h_t = h = c\,n^{-1/(p_0+6)}$, where $p_0 = \max(p+u, 3)$ for the initial OPG step, and $p_0 = \max(d+u, 3)$ during the MAVE refinement loops.

\begin{algorithm}[ht]
\caption{Sequential reduction}\label{algo}
\begin{algorithmic}[1]
\State Compute $K$ and $G=HKH$ with the chosen kernel on $\Omega_Z$ and $\gamma$
\State Form $\W$ and $\Pi_W$
\State For a chosen $(u,\lambda)$, minimize $\widehat J_\lambda(A)$ and estimate $\widehat T$ 
\State Standardize $(\X_i,\widehat T_i)$ and compute $\widehat M$ by partial OPG
\State Choose $\widehat d_x$ and refine $\widehat\B$ by partial MAVE
\State Report $(\widehat\B^\top\X,\widehat T)$
\end{algorithmic}
\end{algorithm}

\subsection{Consistency of Stage 1}\label{subsec:theory1}
{\bf Notation and Setup.}
For any two finite-dimensional subspaces $\mG$ and $\mG'$ with dimension $u$, we define the subspace distance metric $d(\mG,\mG') = \|P_\mG - P_{\mG'}\|_{HS}$. We denote the regularized residual covariance operator as $M_\lambda = \Sigma_\varepsilon + \lambda I$, and define the finite-rank operator $U = \bbeta\Sigma_W\bbeta^*$, which yields the structural decomposition $\Sigma_Z + \lambda I = M_\lambda + U$. Let $(\widehat\Sigma_Z, \widehat\Sigma_\varepsilon, \widehat\bbeta, \widehat\Sigma_W)$ denote the sample estimators of their respective population counterparts, and let $\widehat\Gamma \in \R^{\dim(\mathfrak{M}_Z) \times u}$ be an orthonormal basis matrix whose column span minimizes $\widehat J_\lambda$ over all $u$-dimensional subspaces of the sample range space $\mR = \ran(\Psi) = \ran(\widehat\Sigma_Z)$.

\begin{assumption}\label{ass:S1}
(a) $\kappa$ is uniformly bounded: $\sup_{z \in \Omega_Z} \kappa(z,z) \le \bar\kappa < \infty$.
(b) $\E\|\W\|^4 < \infty$ and $\Sigma_W > \mathbf{0}$.
(c) The envelope subspace $\mE = \mE_{\Sigma_\varepsilon}(\ran\bbeta)$ and $\dim(\mE) = u < \infty$ is known.
(d) $\ran(\bbeta) = \mE$ with $u = \mathrm{rank}(\bbeta)$.
\end{assumption}

Let $\theta_\mE = \lambda_{\min}(\Sigma_Z \vert_\mE) > 0$ denote the minimum eigenvalue of $\Sigma_Z$ compressed to the envelope subspace, let $\sigma_W = \lambda_{\min}(\Sigma_W)$ be the minimum eigenvalue of $\Sigma_W$, and let $s^2 > 0$ represent the smallest non-zero eigenvalue of the operator product $\bbeta^*\bbeta$.

\begin{theorem}\label{thm:stage1}
Under Assumption~\ref{ass:S1} and for a fixed $\lambda > 0$, let $\delta_n$ be the joint convergence rate defined in Lemma~\ref{lem:moments}. Then, on the event $\{\delta_n \le \theta_\mE / 2\}$, whose probability approaches one as $n \to \infty$,
\begin{equation}\label{eq:stage1_bound}
d(\widehat\Gamma, \mE) \le K_\lambda \, \delta_n, \quad \text{where } K_\lambda = \frac{C_\lambda}{c} + \left(\frac{C_2\bar\epsilon^2 + C_\lambda\bar\epsilon}{c}\right)^{1/2}, \quad \bar\epsilon = \frac{\sqrt{2u}}{\theta_\mE},
\end{equation}
and $c, C_2, C_\lambda > 0$ are constants defined in Appendix~\ref{app:stage1}. Consequently, $d(\widehat\Gamma, \mE) = O_P(\lambda^{-3/2}n^{-1/2})$. If $\lambda = \lambda_n \to 0$ such that $n^{1/2}\lambda_n^{3/2} \to \infty$, asymptotic consistency holds with zero asymptotic bias.
\end{theorem}

\subsection{Perturbation of Stage 2}\label{subsec:theory2}
Let $T_i=\Gamma^*(\Phi(\mZ_i)-\mu_Z)$ and $\widehat T_i=\widehat\Gamma^*(\Phi(\mZ_i)-\widehat\mu_Z)$.
\begin{proposition}\label{prop:Tuniform}
Let $\varepsilon_T=\min_{O\in\mathbb O(u)}\max_{i\le n}\|\widehat T_i-O T_i\|$. Then, deterministically, $\varepsilon_T\le2\sqrt{\bar\kappa}\,d(\widehat\Gamma,\mE)+\|\widehat\mu_Z-\mu_Z\|$.  Thus, under Theorem~\ref{thm:stage1}, $\varepsilon_T=O_P(\lambda^{-3/2}n^{-1/2})$. No cross-fitting is needed for this bound even though $\widehat\Gamma$ depends on all $Y_i$.
\end{proposition}

\begin{assumption}\label{ass:D}
(D1) $(\X,T)$ has a density $f$ on a compact $\mathcal D\subset\R^{p+u}$ with $0<f_{\min}\le f\le f_{\max}$ and $f$ Lipschitz; and with $v=(1,\mathbf z)$ and $D_{\mathbf x,h}=\{\mathbf z:\mathbf x+\mathrm{diag}(h_x\mathbf 1_p,h_t\mathbf 1_u)\mathbf z\in\mathcal D\}$,
$\inf_{\mathbf x\in\mathcal D,h\le h_0}\lambda_{\min}\int_{D_{\mathbf x,h}}vv^\top K(\mathbf z)d\mathbf z\ge c_K>0$.
(D2) $K(\mathbf z)=K_x(\mathbf z_x)K_t(\|\mathbf z_t\|)\ge0$ is bounded, Lipschitz, supported in $\{\|\mathbf z_x\|\le1,\|\mathbf z_t\|\le1\}$.
(D3) $h_x\asymp h_t\asymp h\to0$ and $nh^{p+u+2}/\log n\to\infty$.
(D4) $|f_k|\le C_y$; $m_k$ is twice continuously differentiable on $\mathcal D$ with bounded derivatives, $\|\nabla_{\x} m_k\|\le C_m$.
(D5) $\spn(M)=\spn\B(\mG_1)$ has dimension $d_x$ and $g=\lambda_{d_x}(M)>0$ (e.g.\ under the convexity condition in Proposition~\ref{prop:opg}).
\end{assumption}
These assumptions are standard for local linear estimation. (D1) fails when $T$ is concentrated near a lower-dimensional set, as can happen if $\mZ$ depends on a scalar signal and $u>1$. In such scenarios, choose a small $u$ or ridge the local fits.

\begin{theorem}[Consistency of $\widehat\B$ with estimated $T$]\label{thm:stage2}
Under Assumptions~\ref{ass:S1} and~\ref{ass:D}, with $d_x$ known, $\lambda$ fixed (or $\lambda_n$ with $\lambda_n^{-3/2}n^{-1/2}h^{-3}\to0$), and $n^{-1/2}\lambda^{-3/2}h^{-3}\to0$,
\[
\|P_{\widehat\B}-P_{\B(\mG_1)}\|_F\le\frac{2\sqrt{2d_x}}{g}\Big\{\varrho_n+C\,r\,\frac{\varepsilon_T}{h^{3}}\Big\}\quad\text{with probability}\to1,
\]
where $\varrho_n=O_P\big(r\{h+(\log n/(nh^{p+u+2}))^{1/2}\}+n^{-1/2}\big)$ is the oracle error and $\varepsilon_T=O_P(\lambda^{-3/2}n^{-1/2})$. 
\end{theorem}

%==============================================================
\section{Simulation studies}\label{sec:simulations}
%==============================================================
In the numerical studies that follow, the random objects $\mZ$ and $\X$ are associated through a latent signal, which is allowed to be multidimensional.

\subsection{Case 1: Design with a one-dimensional latent signal}\label{subsec:sim_robustness}
Set $(n,p)=(300,10)$, $\X\sim N(\mathbf 0,\Sigma_{XX})$ with $(\Sigma_{XX})_{ij}=0.5^{|i-j|}$, and $\epsilon\sim N(0,1)$ independent of $(\X,\mZ)$. Define $g(u)=u+0.5\sin u$ and the vectors $\mathbf b_1=(1,1,0,\dots,0)^\top/\sqrt{2}$, $\mathbf b_2=(0,\dots,0,1,1)^\top/\sqrt{2}$, and $\mathbf a_0=(0,0,1,-1,0,\dots,0)^\top/\sqrt{2}$.

{\bf Latent signal.} Let $\nu\sim N(0,1)$ be independent of $(\X,\epsilon)$, and let $\xi=\mathbf a_0^\top\X+\nu$. The object $\mZ$ is a function of $\xi$ and independent nuisance variables. Thus, $\mZ$ and $Y$ depend on each other and on $\X$ only through $\xi$, yielding $Y\indep\mZ\mid(\X,\xi)$ and $\mG_Z=\sigma(\xi)$. Because $\xi\mid\X$ is nondegenerate, conditioning on $\X$ does not make $\mZ$ redundant.

{\bf Response models.} Consider the three models:
\begin{align*}
\text{M1: }&Y=1.2\sin(\mathbf b_1^\top\X)+g(\xi)+\epsilon, \\
\text{M2: }&Y=g(0.5\,\mathbf b_1^\top\X+\xi)+\epsilon,\\
\text{M3: }&Y=1.2\sin(\mathbf b_1^\top\X)+\cos(\mathbf b_2^\top\X)\,g(\xi)+\epsilon.
\end{align*}
The ordinary central subspace $\mS_{Y|\X}$ integrates $\xi$ out. For M1, M2, and M3, $\mS_{Y|\X}$ equals $\spn(\mathbf b_1,\mathbf a_0)$, $\spn(0.5\,\mathbf b_1+ \mathbf a_0)$, and $\spn(\mathbf b_1,\mathbf b_2,\mathbf a_0)$, respectively; thus $d_x$ is 1, 1, and 2. In M1 and M3, $\B$ is a proper subspace of $\mS_{Y|\X}$. In M2, the dimensions match but the spans differ, with an angle cosine of $0.5/(0.25+1)^{1/2}=0.447$, which implies that a perfect estimator of $\mS_{Y|\X}$ yields $\Delta = \{2(1-0.447^2)\}^{1/2} \approx 1.265$ from $\spn(\B)$.

{\bf Evaluation.} We use $u=1$ and the true $d_x$ to evaluate $\Delta(\widehat\B,\B)=\|P_{\widehat\B}-P_{\B}\|_F$. $\lambda=10^{-6}$ is used for all estimations. An arbitrary $d$-dimensional subspace of $\R^{10}$ has $\E(\Delta^2)=2d(1-d/p)$, giving baseline benchmarks of $\Delta\approx1.34$ for $d=1$ and $\Delta\approx1.79$ for $d=2$. For non-Euclidean reduction, we report the distance correlation \citep{szekely2007}, $\mathrm{dCor}(\xi,\cdot)$, between $\xi$ and either the first-stage output $\widehat T$ or the competitor principal-component scores using equal dimensions. Competitors evaluated on identical data are: (i) PCA-type reductions of $\mZ$ (FPCA, Wasserstein PCA, MPCA) followed by partial MAVE with the scores as conditioning variables; (ii) MAVE of $Y$ on $\X$ alone, ignoring $\mZ$.

{\it Example 1: Temporal functional predictor.} On a grid of 100 equally spaced points $t\in(0,1)$, let
\[
\mZ_i(t)=e^{\xi_i/2}\sin(6\pi t)+w_{i1}\sin(2\pi t)+w_{i2}\cos(4\pi t),
\]
where $w_{i1}\sim N(0,3^2)$ and $w_{i2}\sim N(0,2^2)$ are independent. The relevant component has smaller variance than the nuisance components. Because the trigonometric terms are orthogonal, $\xi=2\log\{2\int_0^1\mZ(t)\sin(6\pi t)\,dt\}$ is a nonlinear functional of $\mZ$. Under the discrete Fourier distance, the competitor is functional principal component analysis (FPCA) \citep{staniswalis1998nonparametric,yao2005functional}, retaining components explaining at least 90\% of the total variation.

{\it Example 2: Distributional predictor.} Let $\mZ_i=N(\mu_i,\sigma_i^2)$ with quantile function $Q_i(\tau)=\mu_i+\sigma_i\Phi^{-1}(\tau)$, scale $\sigma_i=0.5\,e^{\xi_i/2}$, and independent nuisance location $\mu_i\sim N(0,4)$. Here, $\xi=2\log(2\sigma)$ is a nonlinear functional of $\mZ$. Under the 2-Wasserstein distance, $d_W^2=(\mu-\mu')^2+(\sigma-\sigma')^2$, the metric is dominated by the nuisance $\mu$. The competitor is Wasserstein principal component analysis (WPCA) \citep{verde2015dimension}.

{\it Example 3: Symmetric positive-definite matrix predictor.} Let $\mZ_i=\mathrm{expm}\left(\begin{smallmatrix}\omega_i&\xi_i/2\\ \xi_i/2&\omega_i\end{smallmatrix}\right)$ with independent scale nuisance $\omega_i\sim N(0,1.5^2)$. The eigenvalues are $e^{\omega\pm\xi/2}$, so $\xi=\log(\lambda_+/\lambda_-)$ is a nonlinear functional of $\mZ$. Under the Frobenius distance, the competitor is multilinear principal component analysis (MPCA) \citep{lu2008mpca}, retaining all scores since the matrices are $2\times2$.

\begin{table}[htb!]
\small
\centering
\caption{Simulation results for Examples 1--3 over 100 replications: mean (with standard deviation in parentheses). The best estimator is given in bold.}
\label{tbl:sim_results}
\begin{tabular}{lcccccc}\toprule
 & & \multicolumn{2}{c}{$\mathrm{dCor}(\xi,\cdot)$} & \multicolumn{2}{c}{$\Delta(\widehat\B,\B)$}\\ \cmidrule(lr){3-4}\cmidrule(lr){5-6}
Example & Model & $\widehat T$ & PCA scores & Partial $\spc$ & MAVE on $\X$ only \\ \midrule
1 (Functional)   & M1 & \textbf{0.9120} (0.0207) & 0.1185 (0.0882) & \textbf{0.6514} (0.1722) & 1.2393 (0.0612)\\
                 & M2 & \textbf{0.9123} (0.0215) & 0.1186 (0.0882) & \textbf{0.6319} (0.1648) & 1.2757 (0.0416)\\
                 & M3 & \textbf{0.9005} (0.0256) & 0.1270 (0.0398) & \textbf{1.1947} (0.2656) & 1.5360 (0.1187)\\ \addlinespace
2 (Distribution) & M1 & \textbf{0.9853} (0.0073) & 0.1424 (0.3409) & \textbf{0.3380} (0.0951) & 1.2347 (0.0641)\\
                 & M2 & \textbf{0.9854} (0.0072) & 0.1432 (0.3427) & \textbf{0.3001} (0.0840) & 1.2728 (0.0431)\\
                 & M3 & \textbf{0.9839} (0.0085) & 0.1669 (0.3706) & \textbf{1.0841} (0.2520) & 1.5570 (0.1316)\\ \addlinespace
3 (SPD Matrix)   & M1 & \textbf{0.6813} (0.0691) & 0.6183 (0.0400) & \textbf{0.9950} (0.1332) & 1.2347 (0.0641)\\
                 & M2 & \textbf{0.6880} (0.0570) & 0.6177 (0.0404) & \textbf{0.9725} (0.1365) & 1.2728 (0.0431)\\
                 & M3 & \textbf{0.6603} (0.0818) & 0.5249 (0.0981) & \textbf{1.4499} (0.1783) & 1.5570 (0.1316)\\ \bottomrule
\end{tabular}
\end{table}

{\bf Results.} In Examples~1 and~2, the supervised reduction recovers $\xi$ almost perfectly ($\mathrm{dCor}\ge0.90$), whereas the unsupervised principal components capture high-variance nuisances and retain little signal ($\mathrm{dCor}\le0.17$). For the low-dimensional SPD matrices, principal components are more competitive but still outperformed. For the Euclidean reduction, the partial estimator uniformly achieves smaller $\Delta$ than MAVE on $\X$ alone. In M2, MAVE on $\X$ yields $\Delta\approx 1.27$, matching the theoretical distance of $1.265$ between $\mS_{Y|\X}$ and $\spn(\B)$ because it targets the wrong subspace. Gains are smaller in Example~3 ($\Delta\approx 1.0$ against a baseline of $1.34$ for $d_x=1$; $1.45$ against $1.79$ for $d_x=2$ in M3) due to a weaker signal in $\xi$.

\subsection{Design with a two-dimensional latent signal}\label{subsec:sim_latent2}
This design reverses the dependence between $\X$ and the latent signal from Section~\ref{subsec:sim_robustness}. Here, $\mZ$ is generated first, and $\X$ is drawn from its $\sigma$-field. We fix $(n,p)=(300,10)$, $\epsilon\sim N(0,1)$, $g(u)=u+0.5\sin u$, and keep $\mathbf b_1, \mathbf b_2, \mathbf a_0$ from Section~\ref{subsec:sim_robustness}, adding $\mathbf a_1=(0,0,0,0,1,-1,0,\dots,0)^\top/\sqrt{2}$.

{\bf Predictor generation.} (i) Draw $\bm\xi=(\xi_1,\xi_2)^\top\sim N(\mathbf 0,I_2)$ and independent nuisances. (ii) Build $\mZ$ from $\bm\xi$ and nuisances such that $\sigma(\bm\xi)\subseteq\sigma(\mZ)$ (Examples~4--6 below). (iii) Draw independent $\X_0\sim N(\mathbf 0,\Sig_0)$ with $(\Sig_0)_{ij}=0.5^{|i-j|}$ and set $\X=\X_0+\A\,\bm\xi(\mZ)$ for $\A=[\mathbf a_0,\mathbf a_1]$, yielding $\X\indep\mZ\mid\bm\xi$ and $\Sig_X=\Sig_0+\A\A^\top$. (iv) Generate $Y$ from one of the following response models.

{\bf Response models.} Consider three models:
\begin{align*}
\text{N1: }&Y=1.2\sin(\mathbf b_1^\top\X)+g(\xi_1)+0.5\,e^{\xi_2/2}\,\epsilon,\\
\text{N2: }&Y=1.2\sin(\mathbf b_1^\top\X)+g(\xi_1)+\cos(\mathbf b_2^\top\X)\,g(\xi_2)+\epsilon,\\
\text{N3: }&Y=g\big(0.5\,\mathbf b_1^\top\X+\xi_1+(\mathbf b_1^\top\X)\,\xi_2\big)+\epsilon.
\end{align*}
Here, $Y\indep\mZ\mid(\X,\bm\xi)$ and $\mG_Z=\sigma(\xi_1,\xi_2)$, so $u=2$ for all settings. Given $\bm\xi$, $Y$ depends on $\X$ through $\mathbf b_1^\top\X$ (N1, N3) or $(\mathbf b_1^\top\X,\mathbf b_2^\top\X)$ (N2). Since $\bm\xi\mid\X\sim N(\D^\top\X, \mathbf S)$ with $\D=\Sig_X^{-1}\A = [\bm\delta_1,\bm\delta_2]$ and $\mathbf S=(\I_2 + \A^\top\Sig_0^{-1}\A)^{-1}$, integrating out $\bm\xi$ adds indices $\bm\delta_1^\top\X$ and $\bm\delta_2^\top\X$ to $Y\mid\X$. Thus, $\mS_{Y|\X} = \spn(\mathbf b_1,\bm\delta_1,\bm\delta_2)$ for N1 and N3 ($d_x=1$), and $\mS_{Y|\X} = \spn(\mathbf b_1,\mathbf b_2,\bm\delta_1,\bm\delta_2)$ for N2 ($d_x=2$). In N1, mean-based methods only target $\spn(\mathbf b_1,\bm\delta_1)$. Because $\spn(\B) \subseteq \mS_{Y|\X}$, an exact estimator of $\mS_{Y|\X}$ yields a baseline distance of $\Delta=\{\dim(\mS_{Y|\X})-d_x\}^{1/2}$.

{\it Example 4: Temporal functional predictor.} On the grid $t_i=i/101$ ($i=1,\dots,100$), let $\mZ_i(t)=e^{\xi_{i1}/2}\sin(6\pi t)+e^{\xi_{i2}/2}\sin(10\pi t)+w_{i1}\sin(2\pi t)+w_{i2}\cos(4\pi t)$, where $w_{i1}\sim N(0,3^2)$ and $w_{i2}\sim N(0,2^2)$ are independent. Grid orthogonality of the trigonometric components permits exact recovery via $\xi_k=2\log\{\langle\mZ,s_k\rangle/\|s_k\|^2\}$ for $s_1(t)=\sin(6\pi t)$ and $s_2(t)=\sin(10\pi t)$.

{\it Example 5: Distributional predictor.} Let $\mZ_i$ be a distribution with quantile function $Q_i(\tau)=\mu_i+\sigma_{i1}\,\Phi^{-1}(\tau)+\sigma_{i2}\,\{\Phi^{-1}(\tau)\}^3$, where $\sigma_{i1}=0.5\,e^{\xi_{i1}/2}$, $\sigma_{i2}=0.1\,e^{\xi_{i2}/2}$, and independent nuisance $\mu_i\sim N(0,4)$. Regressing $Q_i-\mu_i$ on $\{\Phi^{-1}(\tau),[\Phi^{-1}(\tau)]^3\}$ over a symmetric grid $\tau\in\{0.01,\dots,0.99\}$ recovers $(\sigma_{i1},\sigma_{i2})$, rendering $\xi_1=2\log(2\sigma_1)$ and $\xi_2=2\log(10\sigma_2)$ as functionals of $\mZ$.

{\it Example 6: SPD matrix predictor.} Let $\mZ_i=e^{\omega_i}\,\mathrm{blockdiag}\{\cosh(\xi_{ik}/2)\I_2+\sinh(\xi_{ik}/2)\mathbf J\}_{k=1,2}\in\mathrm{Sym}^+(4)$, where $\mathbf J=\left(\begin{smallmatrix}0&1\\1&0\end{smallmatrix}\right)$ and independent scale nuisance $\omega_i\sim N(0,1.5^2)$. The eigenvalues are $e^{\omega\pm\xi_k/2}$, making $\xi_k=\log(\lambda_+^{(k)}/\lambda_-^{(k)})$ a functional of $\mZ$.

{\bf Results.} The results are similar to the one-dimensional cases in Examples~1--3. Using $r > 1$ with more complex $\mathbf f(Y)$ did not improve the results in both the simulation and real applications that follow.

\begin{table}[htb!]
\small
\centering
\caption{Simulation results for Examples 4--6 over 100 replications: mean (with standard deviation in parentheses). The best estimator is in bold.}
\label{tbl:sim_results_2d}
\begin{tabular}{lcccccc}\toprule
 & & \multicolumn{2}{c}{$\mathrm{dCor}(\xi,\cdot)$} & \multicolumn{2}{c}{$\Delta(\widehat\B,\B)$}\\ \cmidrule(lr){3-4}\cmidrule(lr){5-6}
Example & Model & $\widehat T$ & PCA scores & Partial $\spc$ & SR on $\X$ only \\ \midrule
4 (Functional)   & N1 & \textbf{0.9101} (0.0793) & 0.1490 (0.0153) & \textbf{0.6993} (0.1660) & 1.2550 (0.0508)\\
                 & N2 & \textbf{0.9016} (0.1107) & 0.1493 (0.0161) & \textbf{1.5238} (0.1374) & 1.7245 (0.1414)\\
                 & N3 & \textbf{0.8955} (0.1253) & 0.1490 (0.0159) & \textbf{0.6054} (0.2026) & 1.2558 (0.0801)\\ \addlinespace
5 (Distribution) & N1 & \textbf{0.7799} (0.0372) & 0.2331 (0.0258) & \textbf{0.5851} (0.1924) & 1.2499 (0.0472)\\
                 & N2 & \textbf{0.7766} (0.0409) & 0.2310 (0.0240) & \textbf{1.5359} (0.1029) & 1.7744 (0.1217)\\
                 & N3 & \textbf{0.7782} (0.0381) & 0.2493 (0.0423) & \textbf{0.6899} (0.2038) & 1.2630 (0.0842)\\ \addlinespace
6 (SPD Matrix)   & N1 & \textbf{0.6202} (0.0797) & 0.4903 (0.0640) & \textbf{1.1681} (0.0785) & 1.2499 (0.0472)\\
                 & N2 & \textbf{0.6149} (0.0854) & 0.3770 (0.0693) & \textbf{1.7316} (0.1041) & 1.7744 (0.1217)\\
                 & N3 & \textbf{0.6153} (0.0805) & 0.4964 (0.0582) & \textbf{1.1681} (0.1413) & 1.2630 (0.0842)\\ \bottomrule
\end{tabular}
\end{table}

%==============================================================
\section{Real applications}\label{sec:application}
%==============================================================
\subsection{Parkinson's disease data}
\begin{figure}[htb!]
\centering
\includegraphics[width=\linewidth]{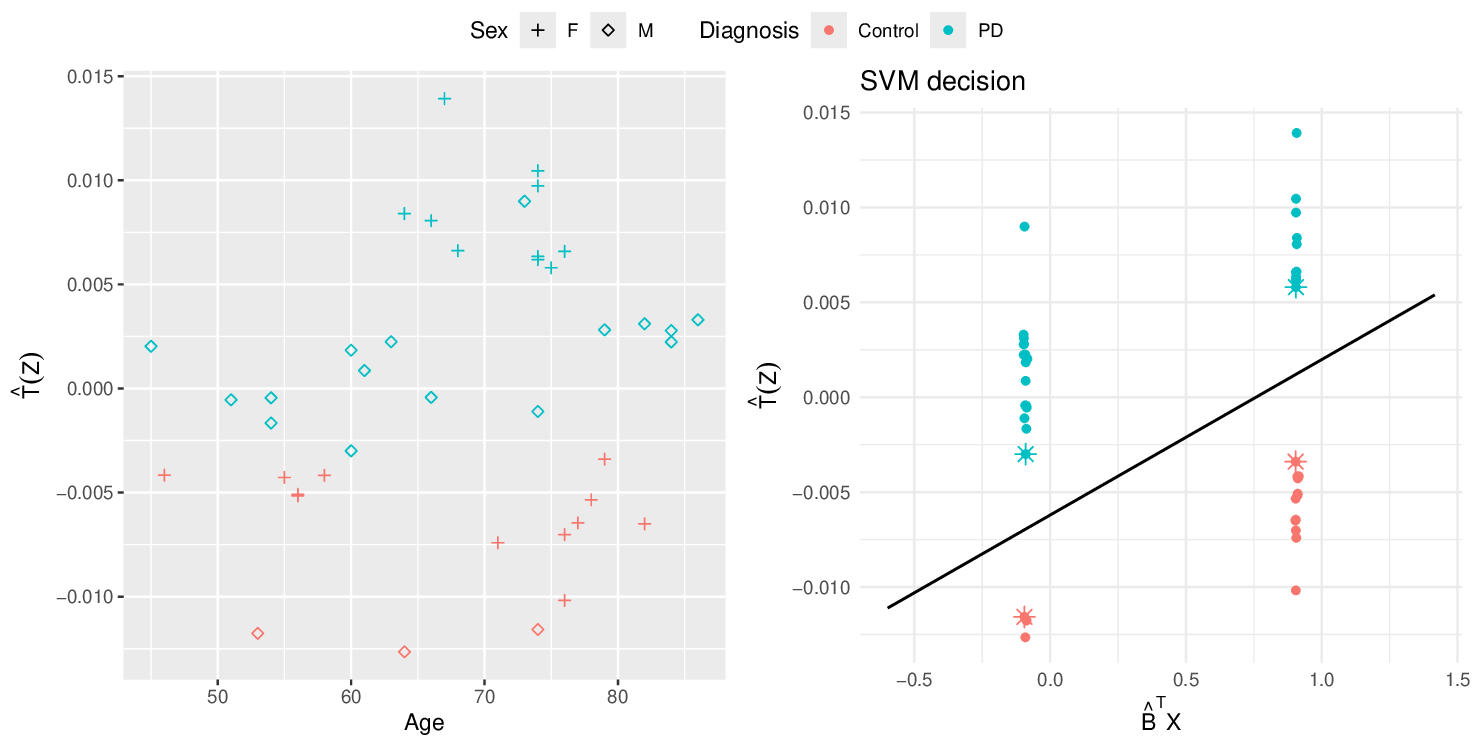}
\caption{Left: in-sample summary plot of the first-stage reduction against age. Right: linear SVM decision boundary on the joint reduction.}
\label{fig:parkinsons}
\end{figure}

We apply the proposed method to the Parkinson's disease data of \cite{badea2017exploring} used in \cite{soale2026metric}. The dataset contains $n=41$ subjects with a binary diagnosis $Y$ (26 Parkinson's disease, 15 controls). The non-Euclidean predictor $\mZ_i \in \mathrm{Sym}(48)$ is a $48\times48$ correlation matrix between regions of interest under the Frobenius distance. The Euclidean predictors are $\X = (\text{age}, \text{sex})$, where sex is a categorical indicator in $\W$ and age is continuous, so $d_x = 1$ and $\B$ is trivial. We set $\mathbf f(Y) = \bm1\{Y=\mathrm{PD}\}$, exercising Stage~1 only to yield the joint reduction $(\text{age}, \text{sex}, \widehat T)$. Figure~\ref{fig:parkinsons} (left) plots the first-stage output against age; the right panel shows a linear support vector machine (SVM) boundary on the joint reduction.

\begin{figure}[htb!]
\centering
\includegraphics{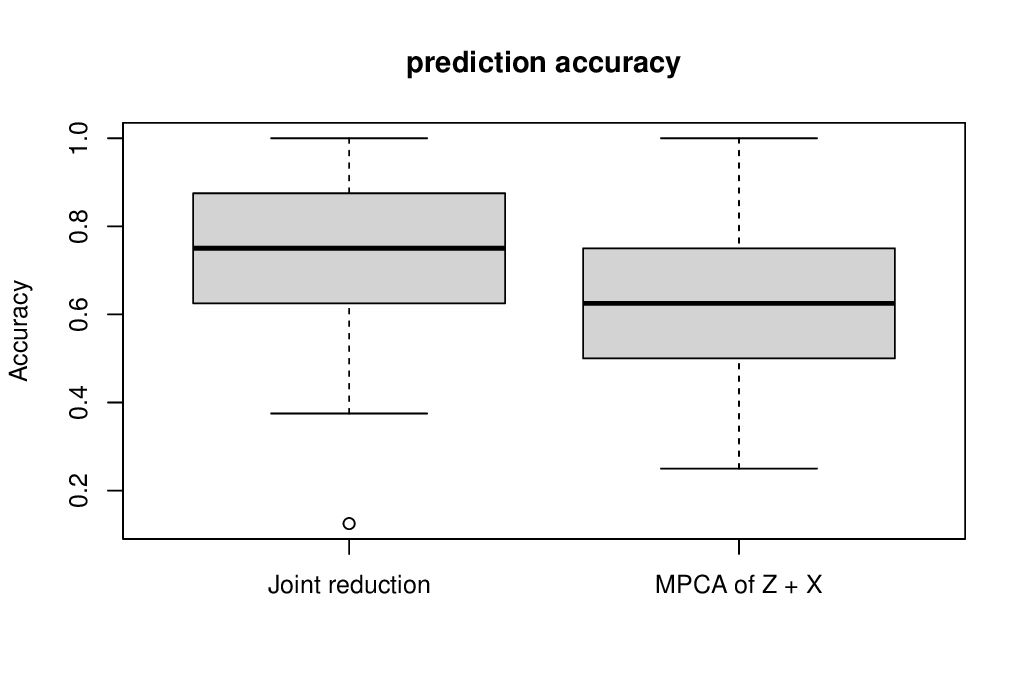}
\caption{Out-of-sample accuracy over 100 random 80/20 partitions.}
\label{fig:pd_pred}
\end{figure}

Out-of-sample performance is evaluated over 100 random 80/20 training/test splits. The reduced space is estimated on the training set, followed by a support vector machine fit on $(\text{age}, \text{sex}, \widehat T)$. As a competitor, multilinear principal component analysis is applied to $\mZ$, and a support vector machine is fit on the scores with age and sex. The proposed joint reduction achieves a median out-of-sample accuracy of 75\% compared to approximately 60\% for the multilinear principal component pipeline, with a narrower spread indicating better consistency (Figure~\ref{fig:pd_pred}).


\subsection{Blood Glucose Data}
We analyze the blood glucose data of \cite{colas2019detrended} for $n=208$ patients. The response $Y$ is basal glycaemia, and $\X$ contains age, gender, body mass index, haemoglobin A1c, follow-up time, and type 2 diabetes status. The non-Euclidean predictor $\mZ$ is a continuous glucose monitoring trajectory in $L^2$, modeled via a Gaussian kernel. We set $\mathbf f(Y)=Y$, and a sequential approach yields optimal dimensions $u=1$ and $d_x=1$.

\begin{figure}[htb!]
\centering
\includegraphics[width=\linewidth]{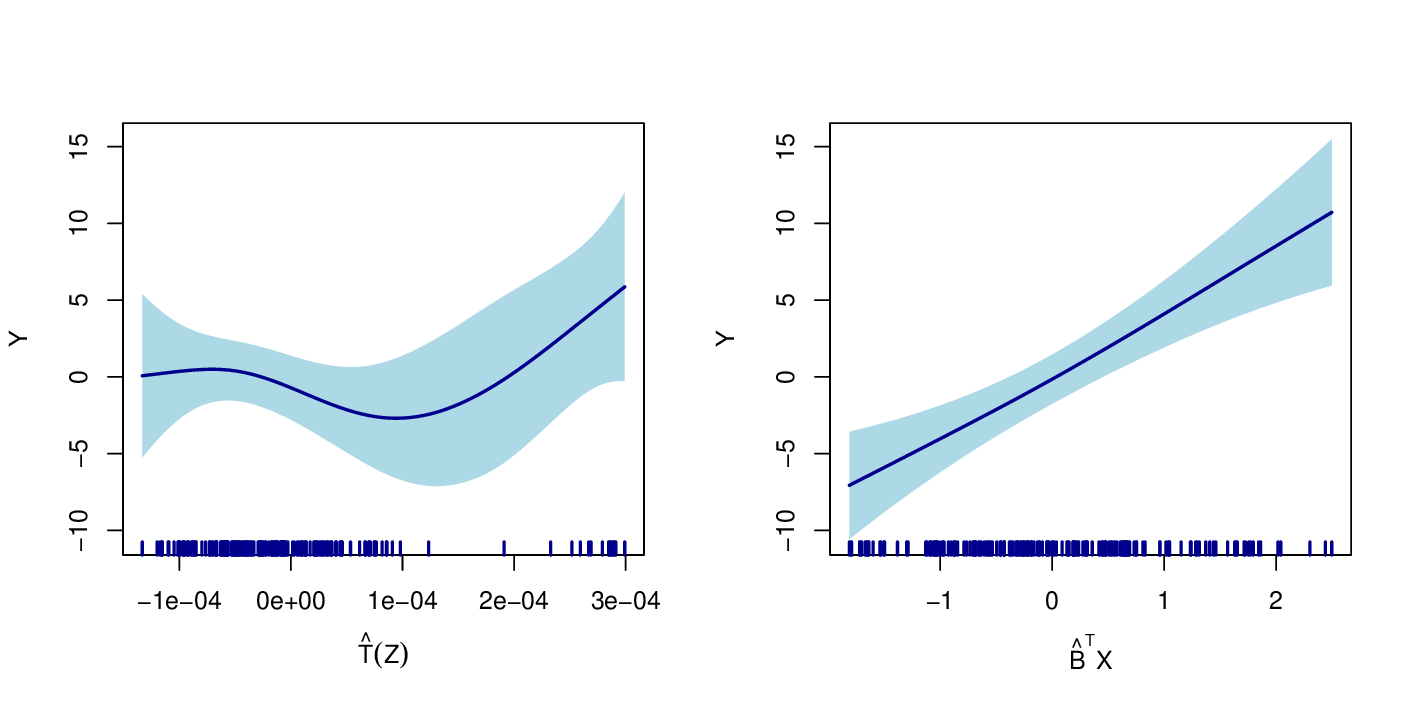}
\caption{Additive model for basal glycaemia on the joint sufficient summaries.}\label{fig:diabetes}
\end{figure}

Using the joint reduction $(\widehat\B^\top\X,\widehat T)$, we fit a generalized additive model with splines. Figure~\ref{fig:diabetes} shows that basal glycaemia depends nonlinearly on the functional summary but linearly on the reduced Euclidean covariates. As shown in Table~\ref{tbl:diabetes}, follow-up time and haemoglobin A1c are the dominant Euclidean predictors followed by gender, whereas body mass index, age, and diabetes status have less influence after controlling for the continuous glucose monitoring profile.

\begin{table}[ht!]
\centering
\caption{Basis estimates $\widehat\B$ based on the partial MAVE}
\label{tbl:diabetes}
\begin{tabular}{llllll}
\hline
Gender  & Age    & BMI    & HbA1c  & FollowUp & T2DM   \\ \hline
-0.3867 & 0.1077 & 0.1016 & 0.5844 & -0.672   & 0.1884 \\ \hline
\end{tabular}
\end{table}

\begin{figure}[htb!]
\centering
\includegraphics{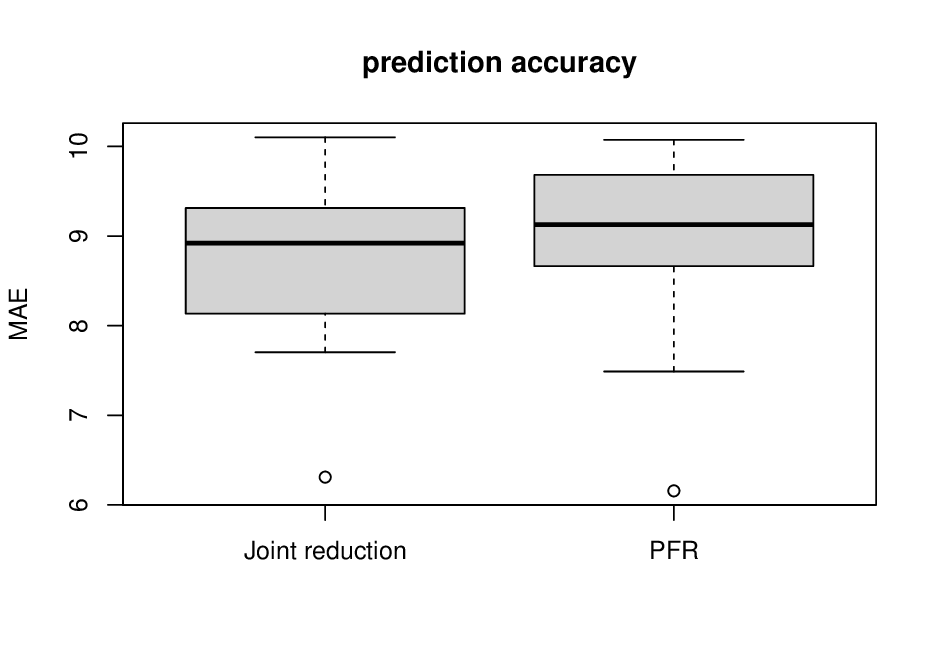}
\caption{Out-of-sample mean absolute error over 100 random 80/20 partitions.}\label{fig:diabetes_pred}
\end{figure}

Out-of-sample performance is evaluated over 100 random 80/20 training/test splits, refitting the entire procedure on each training set. We compare results with penalized functional regression \citep{goldsmith2011penalized}, which accommodates both Euclidean and functional predictors (Figure~\ref{fig:diabetes_pred}). The joint reduction achieves a slightly lower median mean absolute error, suggesting a gain in prediction accuracy.


%==============================================================
\section{Conclusion}\label{sec-conc}
%==============================================================
We introduced a joint sufficient reduction for a Euclidean predictor and a metric-space-valued predictor under arbitrary dependence, estimated sequentially via a kernel-response envelope and partial outer product of gradients or minimum average variance estimation. This target differs from the ordinary central subspace of $Y\mid\X$. Simulations demonstrate that methods ignoring $\mZ$, or reducing it independently of $Y$, can lose crucial information.

Limitations include a first-stage $\sigma$-field that is sufficient but not minimal. Asymptotic theory assumes known dimensions $u$ and $d_x$, requires $\operatorname{ran}\bbeta=\mE$, and relies on local-polynomial conditions that may fail if $T$ is near-degenerate. 

Future directions include establishing consistency for order determination, deriving sharp oracle rates under continuous conditioning, and extending the framework to multiple non-Euclidean predictors.

% \section*{Acknowledgments}
% This was was supported in part by......

%Bibliography
\bibliographystyle{apalike}  
\bibliography{references}  

\appendix

%==============================================================
\section{Proofs}\label{app:proofs}
%==============================================================
\begin{proof}[Proof of Theorem~\ref{thm:joint}]
Let $\mathcal A = \sigma(\X) \vee \sigma(\mZ)$ and $\mathcal C = \sigma(\B^\top\X) \vee \mG \subseteq \mathcal A$. For any bounded measurable function $h$, conditions \eqref{eq:csigmafield} and \eqref{eq:partialcs} imply, respectively,
\[
\E\{h(Y) \mid \mathcal A\} = \E\{h(Y) \mid \sigma(\X) \vee \mG\} = \E\{h(Y) \mid \mathcal C\}.
\]
Because $\mathcal C \subseteq \mathcal A$ and $\E\{h(Y) \mid \mathcal C\}$ is $\mathcal C$-measurable, this identity holds for all $h$. By the definition of conditional independence, it follows that $Y \mathrel{\indep} (\X, \mZ) \mid (\B^\top\X, \mG)$.
\end{proof}

\subsection{Proofs for Stage 1}\label{app:stage1}
Throughout, $\|\cdot\|$ is the operator norm unless stated otherwise, $\|\Phi\|\le\sqrt{\bar\kappa}$ and hence $\|\widetilde\Phi\|\le2\sqrt{\bar\kappa}$, $\|\Sigma_Z\|\le\tr\Sigma_Z\le4\bar\kappa$, and similarly for the sample versions. The operator $\Sigma_\varepsilon\le\Sigma_Z$. Also, $\mE$ reduces $\Sigma_Z$ because it reduces $\Sigma_\varepsilon$ and contains $\ran\bbeta$ (for $x\perp\mE$, $\bbeta^*x=0$ so $\Sigma_Zx=\Sigma_\varepsilon x\in\mE^\perp$).

\begin{proof}[Proof of Proposition~\ref{prop:cond_sufficient}]
(i) By \eqref{eq:UV}, the immaterial part is $V = Q_\mE\varepsilon$, while the material part $U = \bbeta\widetilde\W + P_\mE\varepsilon$ is a measurable function of $(P_\mE\varepsilon, \widetilde\W)$. Thus, $Q_\mE\varepsilon \indep (P_\mE\varepsilon, \widetilde\W)$ implies $V \indep (U, \widetilde\W)$, which by weak union yields $V \indep \widetilde\W \mid U$.

(ii) Covariance orthogonality $P_\mE\Sigma_\varepsilon Q_\mE = \mathbf{0}$ implies independence between the jointly Gaussian elements $P_\mE\varepsilon$ and $Q_\mE\varepsilon$. Because both are independent of $\widetilde\W$, condition (i) holds. 
\end{proof}

\begin{proof}[Proof of Theorem~\ref{thm:suff}]
Because $\widetilde\Phi = U + V$ and $U$ is $\sigma(U)$-measurable, Assumption~\ref{ass:cond_indep} implies $(U, V) \indep \W \mid U$. As $\widetilde\Phi$ is a measurable function of $(U, V)$, the relation $\widetilde\Phi \indep \W \mid U$ follows. Since $\Gamma$ is an isometry onto $\mE$, we have $\Gamma^*\Gamma = I_u$ and $P_\mE = \Gamma\Gamma^*$. Thus, $U = P_\mE\widetilde\Phi = \Gamma T$ and $T = \Gamma^*U$, establishing $\sigma(U) = \sigma(T)$. This mutual measurability and centering transformations yield $\Phi(\mZ) \indep \W \mid T$.

Because $f_1$ is an injective Borel map with a standard Borel inverse, $(\X, Y)$ is a measurable function of $\W$. By decomposition, $\Phi(\mZ) \indep (\X, Y) \mid T$. Finally, since $T$ is a measurable function of $\Phi(\mZ)$, Lemma~\ref{lem:seq} establishes \eqref{eq:response-envelope-sufficiency}.
\end{proof}

\begin{lemma}\label{lem:moments}
Under Assumptions~\ref{ass:sparse} and~\ref{ass:cond_indep}, the sample moments satisfy
$\delta_n:=\max\{\|\widehat\Sigma_Z-\Sigma_Z\|_{HS},\|\widehat\Sigma_\varepsilon-\Sigma_\varepsilon\|_{HS},\|\widehat\bbeta-\bbeta\|_{HS},\|\widehat\mu_Z-\mu_Z\|\}=O_P(n^{-1/2})$.
\end{lemma}

\begin{proof}[Proof of Lemma~\ref{lem:moments}]
Because $\widetilde\Phi_i$ and $\widetilde\Phi_i \otimes \widetilde\Phi_i$ have bounded support in $\mathfrak M_Z$ and $HS(\mathfrak M_Z)$, respectively, their population variances are finite. Chebyshev's inequality for Hilbert spaces implies $\| \widehat\mu_Z - \mu_Z\|= O_P(n^{-1/2})$ and $\| \widehat\Sigma_Z - \Sigma_Z\|_{\mathrm{HS}} = O_P(n^{-1/2})$. Standard multivariate theory for the finite-dimensional $\W$ with $\Sigma_W > 0$ yields $\|\widehat\Sigma_W^{-1}-\Sigma_W^{-1}\|_F=O_P(n^{-1/2})$. Under $\E\|\W\|^4<\infty$, bounded support implies $\E\|\widetilde\Phi\|^2\|\widetilde\W\|^2 \le 4\bar\kappa\,\E\|\widetilde\W\|^2 < \infty$, so $\|\widehat\Sigma_{ZW}-\Sigma_{ZW}\|_{\mathrm{HS}}=O_P(n^{-1/2})$.

For the regression operator, decomposing $\widehat\bbeta-\bbeta$ as $(\widehat\Sigma_{ZW}-\Sigma_{ZW})\widehat\Sigma_W^{-1}+\Sigma_{ZW}(\widehat\Sigma_W^{-1}-\Sigma_W^{-1})$ and applying the triangle inequality with the submultiplicativity of mixed norms yields
\[
\|\widehat\bbeta-\bbeta\|_{\mathrm{HS}} \leq  \|\widehat\Sigma_{ZW}-\Sigma_{ZW}\|_{\mathrm{HS}}\|\widehat\Sigma_W^{-1}\|_F + \|\Sigma_{ZW}\|_{\mathrm{HS}}\|\widehat\Sigma_W^{-1}-\Sigma_W^{-1}\|_F = O_P(n^{-1/2}).
\]
Telescoping the terms in $\widehat\bbeta\widehat\Sigma_W\widehat\bbeta^* = \widehat\Sigma_{ZW}\widehat\Sigma_W^{-1}\widehat\Sigma_{WZ}$ similarly establishes $\|\widehat\Sigma_{ZW}\widehat\Sigma_W^{-1}\widehat\Sigma_{WZ}-\Sigma_{ZW}\Sigma_W^{-1}\Sigma_{WZ}\|_{\mathrm{HS}}=O_P(n^{-1/2})$. Finally, the result follows by applying the triangle inequality to the residual covariance decomposition $\widehat\Sigma_\varepsilon-\Sigma_\varepsilon=(\widehat\Sigma_Z-\Sigma_Z)-(\widehat\Sigma_{ZW}\widehat\Sigma_W^{-1}\widehat\Sigma_{WZ}-\Sigma_{ZW}\Sigma_W^{-1}\Sigma_{WZ})$, which boundedly yields $\|\widehat\Sigma_\varepsilon-\Sigma_\varepsilon\|_{\mathrm{HS}} = O_P(n^{-1/2})$. 
\end{proof}

\begin{lemma}\label{lem:gap}
Let $H$ be a bounded and boundedly invertible linear operator on a Hilbert space, and let $\mG$ and $\mG'$ be subspaces of the same finite dimension $u$. Then, the orthogonal projection operators satisfy:
\[
\|P_{H\mG} - P_{H\mG'}\|_{\mathrm{op}} \le \kappa(H) \|P_\mG - P_{\mG'}\|_{\mathrm{op}},
\]
where $\kappa(H) = \|H\| \|H^{-1}\|$ is the condition number of $H$.
\end{lemma}

Lemma~\ref{lem:gap} is a standard results from classical operator perturbation literature, see  \cite{kato1995perturbation}

\begin{lemma}\label{lem:lip}
For a $u$-dimensional subspace $\mG$ with orthonormal basis operator $\Gamma: \R^u \to \mathfrak M_Z$, let $J_\lambda(\mG)=\log|\Gamma^*M\Gamma|+\log|\Gamma^*A^{-1}\Gamma|$, where $M = \Sigma_\varepsilon + \lambda I$ and $A = \Sigma_Z + \lambda I$. Let $\Delta(\mG)=\widehat J_\lambda(\mG)-J_\lambda(\mG)$. Then on $\{\delta_n\le1\}$, all $u$-dimensional subspaces $\mG, \mG'$ satisfy $|\Delta(\mG)-\Delta(\mG')|\le C_\lambda\delta_nd(\mG,\mG')$, where $C_\lambda=4u\kappa/\lambda$ and $\kappa=\{(4\bar\kappa+\lambda)/\lambda\}^{1/2}$. As $\lambda \to 0$, $C_\lambda=O(u\lambda^{-3/2})$.
\end{lemma}

\begin{proof}[Proof of Lemma~\ref{lem:lip}]
Let $\widehat M = \widehat\Sigma_\varepsilon +\lambda I$ and $\widehat A=\widehat\Sigma_Z+\lambda I$. Define the linear interpolations $M_t=(1-t)M+t\widehat M$ and $A_t=(1-t)A+t\widehat A$ for $t \in [0,1]$. Differentiating the log-determinants yields $\tfrac{d}{dt}\log|\Gamma^*M_t\Gamma|=\tr\{\Delta M\,\Pi^M_t(\mG)\}$ and $\tfrac{d}{dt}\log|\Gamma^*A_t^{-1}\Gamma|=-\tr\{\Delta A\,\Pi^A_t(\mG)\}$, where $\Delta M = \widehat\Sigma_\varepsilon - \Sigma_\varepsilon$, $\Delta A = \widehat\Sigma_Z - \Sigma_Z$, $\Pi^M_t(\mG)=M_t^{-1/2}P_{M_t^{1/2}\mG}M_t^{-1/2}$, and $\Pi^A_t(\mG)=A_t^{-1/2}P_{A_t^{-1/2}\mG}A_t^{-1/2}$. By the fundamental theorem of calculus,
\[
\Delta(\mG)-\Delta(\mG')=\int_0^1 \tr\big\{\Delta M(\Pi^M_t(\mG)-\Pi^M_t(\mG'))\big\}\,dt - \int_0^1 \tr\big\{\Delta A(\Pi^A_t(\mG)-\Pi^A_t(\mG'))\big\}\,dt,
\]
where $\|\Delta M\|_{\mathrm{op}}, \|\Delta A\|_{\mathrm{op}} \le \delta_n$. Because the covariance operators are positive semi-definite, $\|M_t^{-1}\_{\mathrm{op}}, \|A_t^{-1}\_{\mathrm{op}} \le \lambda^{-1}$.

Since $\Pi^M_t(\mG) - \Pi^M_t(\mG')$ has rank at most $2u$, H\"{o}lder's inequality for operators yields $|\tr\{\Delta M(\Pi^M_t(\mG)-\Pi^M_t(\mG'))\}| \le 2u \|\Delta M\|_{\mathrm{op}} \|\Pi^M_t(\mG)-\Pi^M_t(\mG')\|_{\mathrm{op}}$. Applying Lemma~\ref{lem:gap} bounds the operator norm distance by
\[
\|\Pi^M_t(\mG)-\Pi^M_t(\mG')\|_{\mathrm{op}} \le \|M_t^{-1}\|_{\mathrm{op}} \|P_{M_t^{1/2}\mG} - P_{M_t^{1/2}\mG'}\|_{\mathrm{op}} \le \lambda^{-1} \kappa d(\mG, \mG'),
\]
where $\kappa \le \{(4\bar\kappa+\lambda)/\lambda\}^{1/2}$. Combining these steps, both integrands are bounded by $2u\lambda^{-1}\kappa\delta_n d(\mG, \mG')$. Integrating over $t \in [0,1]$ yields the Lipschitz coefficient $C_\lambda = 4u\kappa/\lambda$. 
\end{proof}

\begin{lemma}\label{lem:rep}
Let $\mR = \operatorname{ran}(\widehat\Sigma_Z)$. Then $\|(I-P_{\mR})P_\mE\|_{\mathrm{op}} \le \theta_\mE^{-1}\|\widehat\Sigma_Z - \Sigma_Z\|_{\mathrm{op}}$. Furthermore, on $\{\delta_n \le \theta_\mE / 2\}$, the projected subspace $\widetilde\mE = P_{\mR}\mE \subseteq \mR$ satisfies $\dim(\widetilde\mE) = u$ and $d(\widetilde\mE, \mE) \le \bar\epsilon \, \delta_n$, where $\bar\epsilon = 2\sqrt{2u}/\theta_\mE$.
\end{lemma}

\begin{proof}[Proof of Lemma~\ref{lem:rep}]
Because $\mE$ reduces $\Sigma_\varepsilon$ and contains $\operatorname{ran}(\bbeta)$, it reduces $\Sigma_Z = \Sigma_\varepsilon + \bbeta\Sigma_W\bbeta^*$, meaning $\Sigma_Z \mE \subseteq \mE$. Under $\theta_\mE > 0$, any $e \in \mE$ equals $\Sigma_Z f$ for a unique $f \in \mE$ with $\|f\| \le \|e\| / \theta_\mE$. Since $(I-P_{\mR})\widehat\Sigma_Z = \mathbf{0}$, the range escape error decomposes as $(I-P_{\mR})e = (I-P_{\mR})(\Sigma_Z - \widehat\Sigma_Z)f$. Supremum over unit vectors $\|e\| = 1$ establishes the first bound: $\|(I-P_{\mR})P_\mE\|_{\mathrm{op}} \le \theta_\mE^{-1}\|\widehat\Sigma_Z - \Sigma_Z\|_{\mathrm{op}}$.

Let $\epsilon_0 = \theta_\mE^{-1}\|\widehat\Sigma_Z - \Sigma_Z\|_{\mathrm{op}}$. On $\{\delta_n \le \theta_\mE / 2\}$, we have $\epsilon_0 \le 1/2$. By the Pythagorean identity, $\|P_{\mR} e\|^2 = \|e\|^2 - \|(I-P_{\mR})e\|^2 \ge (1 - \epsilon_0^2)\|e\|^2 \ge (3/4)\|e\|^2$. Thus, $P_{\mR}$ is injective on $\mE$, ensuring $\dim(\widetilde\mE) = \dim(\mE) = u$. For any unit $x = P_{\mR} e / \|P_{\mR} e\| \in \widetilde\mE$, its directed distance satisfies
\[
\operatorname{dist}(x, \mE) = \|(I-P_\mE)x\| = \frac{\|(I-P_\mE)(I-P_{\mR})e\|}{\|P_{\mR} e\|} \le \frac{\epsilon_0}{(1-\epsilon_0^2)^{1/2}} \le 2\epsilon_0.
\]
Because $\dim(\widetilde\mE) = \dim(\mE) = u$, the projection operator norm difference equals this directed distance, yielding $\|P_{\widetilde\mE} - P_\mE\|_{\mathrm{op}} \le 2\epsilon_0$. Since $\operatorname{rank}(P_{\widetilde\mE} - P_\mE) \le 2u$, its Hilbert--Schmidt norm satisfies
\[
d(\widetilde\mE, \mE) = \|P_{\widetilde\mE} - P_\mE\|_{\mathrm{HS}} \le (2u)^{1/2} \|P_{\widetilde\mE} - P_\mE\|_{\mathrm{op}} \le 2(2u)^{1/2}\epsilon_0 \le \frac{2(2u)^{1/2}}{\theta_\mE}\delta_n = \bar\epsilon\,\delta_n.
\] 
\end{proof}

\begin{proof}[Proof of Theorem~\ref{thm:stage1}]
On the event $\{\delta_n \le \min(1, \theta_\mE / 2)\}$, whose probability approaches one by Lemma~\ref{lem:moments}, let $d = d(\widehat\Gamma, \mE)$ and $\epsilon = d(\widetilde\mE, \mE) \le \bar\epsilon \delta_n$ via Lemma~\ref{lem:rep}. Under Assumption~\ref{ass:S1}(d), quadratic growth satisfies $c d^2 \le J_\lambda(\widehat\Gamma) - J_\lambda(\mE) = \{ \widehat J_\lambda(\widehat\Gamma) - \widehat J_\lambda(\mE) \} + \Delta(\mE) - \Delta(\widehat\Gamma)$ for a constant $c > 0$. Since $\widehat\Gamma$ minimizes $\widehat J_\lambda$ over the sample range $\mR$ and $\widetilde\mE \subseteq \mR$, we have $\widehat J_\lambda(\widehat\Gamma) \le \widehat J_\lambda(\widetilde\mE)$, implying $\widehat J_\lambda(\widehat\Gamma) - \widehat J_\lambda(\mE) \le \{ J_\lambda(\widetilde\mE) - J_\lambda(\mE) \} + \{ \Delta(\widetilde\mE) - \Delta(\mE) \big\}$.

A second-order expansion yields $J_\lambda(\widetilde\mE) - J_\lambda(\mE) \le C_2 \epsilon^2$. Combining this with the uniform Lipschitz continuity from Lemma~\ref{lem:lip} gives $\widehat J_\lambda(\widehat\Gamma) - \widehat J_\lambda(\mE) \le C_2 \epsilon^2 + C_\lambda \delta_n \epsilon$ and $|\Delta(\mE) - \Delta(\widehat\Gamma)| \le C_\lambda \delta_n d$. These bounds combine to produce the quadratic inequality
\[
d^2 \le (C_\lambda/c)\delta_n d + \{(C_2\bar\epsilon^2 + C_\lambda\bar\epsilon)/c\}\delta_n^2.
\]
Solving via $x^2 \le ax + b \implies x \le a + b^{1/2}$ yields $d \le K_\lambda \delta_n$. As $\lambda \to 0$, the coefficient $C_\lambda = O(\lambda^{-3/2})$ dominates both $\sqrt{C_2} = O(\lambda^{-1})$ and $\sqrt{C_\lambda} = O(\lambda^{-3/4})$, so $K_\lambda = O(\lambda^{-3/2})$. Substituting $\delta_n = O_P(n^{-1/2})$ completes the proof. 
\end{proof}

\subsection{Proofs for Stage 2}\label{app:stage2}
\begin{proof}[Proof of Proposition~\ref{prop:Tuniform}]
For an arbitrary orthogonal matrix $O \in \mathbb O(u)$, decompose the pointwise discrepancy between $T_i = \Gamma^*(\Phi_i - \mu_Z)$ and $\widehat T_i = \widehat\Gamma^*(\Phi_i - \widehat\mu_Z)$ as
\[
\widehat T_i - OT_i = \big(\widehat\Gamma - \Gamma O^\top\big)^*(\Phi_i - \mu_Z) + \widehat\Gamma^*(\mu_Z - \widehat\mu_Z).
\]
Because centered features are bounded as $\|\Phi_i - \mu_Z\| \le 2\sqrt{\bar\kappa}$ almost surely, let $\theta_1, \dots, \theta_u$ be the principal angles between $\operatorname{ran}(\widehat\Gamma)$ and $\operatorname{ran}(\Gamma) = \mE$. Procrustes alignment properties \citep{kato1995perturbation} imply
\[
\min_{O \in \mathbb O(u)} \|\widehat\Gamma - \Gamma O^\top\|_{\mathrm{op}}^2 \le \min_{O \in \mathbb O(u)} \|\widehat\Gamma - \Gamma O^\top\|_F^2 = 2\sum_{l=1}^u (1 - \cos\theta_l) \le \|P_{\widehat\Gamma} - P_\Gamma\|_F^2 = d(\widehat\Gamma, \mE)^2,
\]
since $1 - \cos\theta \le \sin^2\theta$. Choosing the optimal Procrustes rotation $O^*$ and applying the triangle inequality yields
\[
\|\widehat T_i - O^*T_i\| \le \|\widehat\Gamma - \Gamma (O^*)^\top\|_{\mathrm{op}} \|\Phi_i - \mu_Z\| + \|\widehat\Gamma\|_{\mathrm{op}} \|\mu_Z - \widehat\mu_Z\| \le 2\sqrt{\bar\kappa}\,d(\widehat\Gamma, \mE) + \|\mu_Z - \widehat\mu_Z\|,
\]
where $\|\widehat\Gamma\|_{\mathrm{op}} = 1$. By Lemma~\ref{lem:moments} and Theorem~\ref{thm:stage1}, $\|\mu_Z - \widehat\mu_Z\| = O_P(n^{-1/2})$ and $d(\widehat\Gamma, \mE) = O_P(n^{-1/2}\lambda^{-3/2})$, ensuring the bound holds uniformly across all $i \le n$. 
\end{proof}

{\it Local fit notation.} Fix a focal observation index $j \le n$ and $k \le r$, and let $y_i = f_k(Y_i)$ and $d = p + u$. Assume bandwidths satisfy $h_x \asymp h_t \asymp h$. Define the product kernel weights as $K_{ij} = K_x (\|\X_i - \X_j\|/h_x ) K_t (\|T_i - T_j\|/h_t )$ and $w_{ij} = K_{ij}/\sum_{l=1}^n K_{lj}$ for isotropic kernels $K_x, K_t$. Let $\mathbf{z}_i = ( (\X_i - \X_j)/h_x, (T_i - T_j)/h_t )^\top$ and $\mathbf{v}_i = ( 1, \mathbf{z}_i )^\top$. The local linear solution is $\boldsymbol\theta = S^{-1}\mathbf{s} \in \R^{1+d}$, where $S = \sum_{i=1}^n w_{ij} \mathbf{v}_i \mathbf{v}_i^\top$ and $\mathbf{s} = \sum_{i=1}^n w_{ij} \mathbf{v}_i y_i$.

The local estimated gradient with respect to $\X$ is $\mathbf{b} = h_x^{-1} \boldsymbol\theta_{\X}$, where $\boldsymbol\theta_{\X} \in \R^p$ is the sub-vector corresponding to $\X$. Tilded quantities ($\widetilde{K}_{ij}$, $\widetilde{w}_{ij}$, $\widetilde{S}$, $\widetilde{\mathbf{s}}$, and $\widetilde{\mathbf{b}}$) denote sample estimators evaluated by substituting the estimated scores $\widehat{T}_i$ for the optimally rotated population coordinates $O^*T_i$. Because $K_t$ is radial, $\mathbf{b}$ is invariant to the choice of rotation matrix $O \in \mathbb{O}(u)$. Finally, define the normalized score perturbation ratio as $\tau = 2\varepsilon_T / h_t$, where $\max_i \|\widehat{T}_i - O^*T_i\| \le \varepsilon_T$.

\begin{lemma}\label{lem:pert}
Suppose that $\lambda_{\min}(S) \ge 2c_0$, $\tau \le \tau_0$, $|y_i| \le C_y$, and that the localized coordinates satisfy $\|\widetilde{\mathbf z}_i\| \le \rho_0$, $\|\mathbf z_i\| \le \rho_0$, and $\|\widetilde{\mathbf z}_i - \mathbf z_i\| \le \tau$ for all indices $i$ with positive weights. Let $\sum_{i=1}^n |\widetilde w_{ij} - w_{ij}| \le C_w \tau$ and $V = 1 + \rho_0^2$. If $(C_wV + 2\sqrt V)\tau_0 \le c_0$, then $\lambda_{\min}(\widetilde S) \ge c_0$ and 
\begin{equation}\label{eq:theta_bounds}
\|\widetilde{\boldsymbol\theta} - \boldsymbol\theta\| \le C_\theta \tau, \qquad \|\boldsymbol\theta\| \le \frac{C_y\sqrt V}{2c_0},
\end{equation}
where $C_\theta = c_0^{-1}C_y(C_w\sqrt V + 1) + 2^{-1}c_0^{-2}C_y\sqrt V(C_wV + 2\sqrt V)$. Consequently, $\|\widetilde{\mathbf b} - \mathbf b\| \le C_\theta\tau / h_x$, $\|\mathbf b\| \le C_y\sqrt V / (2c_0h_x)$, and the matrix perturbation satisfies
\begin{equation}\label{eq:matrix_opg_bound}
\|\widehat M(\widehat T) - \widehat M(\T)\|_{\mathrm{op}} \le \frac{C \, r \, \tau}{h_x^2} = \frac{2 C r \varepsilon_T}{h_x^2 h_t}.
\end{equation}
\end{lemma}

\begin{proof}[Proof of Lemma~\ref{lem:pert}]
Applying the triangle inequality to the design matrix difference yields
\[
\|\widetilde S - S\|_{\mathrm{op}} \le \sum_{i=1}^n |\widetilde w_{ij} - w_{ij}| \|\widetilde{\mathbf v}_i\|^2 + \sum_{i=1}^n w_{ij} \|\widetilde{\mathbf v}_i\widetilde{\mathbf v}_i^\top - \mathbf v_i\mathbf v_i^\top\|_{\mathrm{op}} \le (C_wV + 2\sqrt V)\tau \le c_0.
\]
By Weyl's theorem, this guarantees $\lambda_{\min}(\widetilde S) \ge \lambda_{\min}(S) - \|\widetilde S - S\|_{\mathrm{op}} \ge 2c_0 - c_0 = c_0$. Similarly, bounding the response vector gives $\|\widetilde{\mathbf s} - \mathbf s\| \le C_y(C_w\sqrt V + 1)\tau$. Leveraging the matrix identity
\[
\widetilde{\boldsymbol\theta} - \boldsymbol\theta = \widetilde S^{-1}(\widetilde{\mathbf s} - \mathbf s) + \widetilde S^{-1}(S - \widetilde S)S^{-1}\mathbf s
\]
with $\|S^{-1}\|_{\mathrm{op}} \le (2c_0)^{-1}$, $\|\widetilde S^{-1}\|_{\mathrm{op}} \le c_0^{-1}$, and $\|S^{-1}\mathbf s\| \le C_y\sqrt V / (2c_0)$, the triangle inequality establishes \eqref{eq:theta_bounds} under $C_\theta$. Extracting the sub-blocks yields the gradient bounds. Averaging the rank-one expansion $\|\widetilde{\mathbf b}\widetilde{\mathbf b}^\top - \mathbf b\mathbf b^\top\|_{\mathrm{op}} \le \|\widetilde{\mathbf b} - \mathbf b\| (2\|\mathbf b\| + \|\widetilde{\mathbf b} - \mathbf b\|)$ over $j \le n$ and summing across $k \le r$ establishes \eqref{eq:matrix_opg_bound}. 
\end{proof}

\begin{lemma}\label{lem:design}
Under Assumptions~\ref{ass:sparse} and~\ref{ass:cond_indep}, with probability approaching one, the following hold uniformly over $j\le n$ and $\tau\le1$: 
(i) $\lambda_{\min}(S_j)\ge2c_0$ for a constant $c_0>0$; 
(ii) $\#\{i:K_{ij}>0\text{ or }\widetilde K_{ij}>0\}\le C_1nh^d$; 
(iii) $\sum_{i=1}^n K_{ij}\ge C_2nh^d$. 
Consequently, the hypotheses of Lemma~\ref{lem:pert} hold uniformly with $\rho_0=\sqrt5$ and $C_w=4L_tC_1/C_2$, where $L_t$ is the Lipschitz constant of $K_t$ and $\|K_x\|_\infty, \|K_t\|_\infty\le1$.
\end{lemma}

\noindent Lemma~\ref{lem:design} follows from uniform consistency bounds under Lipschitz kernels \citep{einmahl2005}.

\begin{lemma}\label{lem:oracle}
Under Assumptions~\ref{ass:sparse}, \ref{ass:cond_indep}, and the smoothness condition (D4),
\[
\|\widehat M(T) - M\|_{\mathrm{op}} = O_P\left( r\left\{ h + \left( \frac{\log n}{nh^{d+2}} \right)^{1/2} \right\} + r n^{-1/2} \right) =: \rho_n.
\]
\end{lemma}
Lemma~\ref{lem:oracle} directly mirrors the standard empirical process framework of \cite{einmahl2005} and the gradient extraction rate matrices found in \cite{xia2002adaptive}.

\begin{proof}[Proof of Theorem~\ref{thm:stage2}]
Let $\widehat\B \in \R^{p \times d_x}$ be formed by the leading $d_x$ eigenvectors of $\widehat M = \widehat M(\widehat T)$. By Proposition~\ref{prop:Tuniform} and Theorem~\ref{thm:stage1}, $\varepsilon_T = \max_i \|\widehat T_i - O^*T_i\| = O_P(\lambda^{-3/2}n^{-1/2})$. Under $h_x \asymp h_t \asymp h$, the perturbation scale satisfies $\tau = 2\varepsilon_T / h_t = O_P(\varepsilon_T / h)$. The condition $\varepsilon_T / h^3 \to 0$ ensures $\tau \to 0$, satisfying the uniform design constraints of Lemma~\ref{lem:design}. On this event, Lemma~\ref{lem:pert} yields
\[
\big\|\widehat M(\widehat T) - \widehat M(T)\big\|_{\mathrm{op}} \le \frac{C \, r \, \varepsilon_T}{h_x^2 h_t} \le \frac{C \, r \, \varepsilon_T}{h^3}.
\]
Because Lemma~\ref{lem:oracle} guarantees $\big\|\widehat M(T) - M\big\|_{\mathrm{op}} \le \rho_n$, the triangle inequality bounds the total matrix perturbation by $\eta := \|\widehat M - M\|_{\mathrm{op}} \le \rho_n + Cr\varepsilon_T/h^3$.

Under condition (D5), $M$ has rank $d_x$ with its smallest non-zero eigenvalue $\lambda_{d_x}(M) = g > 0$, establishing an eigengap of $g$. By the Davis--Kahan theorem \citep[Thm.~2]{yu2015useful}, the canonical angles satisfy $\|\sin\Theta(\widehat\B, \B)\|_F \le 2\sqrt{d_x}\,\eta / g$. Since $d_x$ is finite and fixed, $\|P_{\widehat\B} - P_\B\|_F = \sqrt{2}\|\sin\Theta(\widehat\B, \B)\|_F$. Substituting the error bounds yields the final rate:
\[
d(\widehat\B, \B) = \|P_{\widehat\B} - P_\B\|_F \le \frac{2\sqrt{2d_x}}{g} \eta = O_P\left( \rho_n + \frac{r\,\varepsilon_T}{h^3} \right). 
\] 
\end{proof}

\end{document}